\documentclass[11pt]{article}
\usepackage{hyperref}
\usepackage{amsmath,amssymb,amsfonts}
\usepackage{mathrsfs}
\usepackage{graphicx}
\usepackage{hyperref,url}
\usepackage{amssymb}
\usepackage{mathrsfs}
\usepackage{amsthm}
\usepackage{graphicx}
\usepackage{color}
\usepackage{multirow}
\usepackage{amsmath,amssymb,amsfonts}
\usepackage{amsthm}
\usepackage{mathrsfs}
\usepackage[title]{appendix}
\usepackage{xcolor}
\usepackage{textcomp}
\usepackage{manyfoot}
\usepackage{booktabs}
\usepackage{listings}

\begin{document}

\title{\large \bf The $\alpha$-Attractor E-Model in Warm Inflation: Observational Viability from Planck 2018}

\author{\normalsize{\sc Bhargabi Saha}\footnote{{\tt bhargabi@iitg.ac.in}}\ \   and {\sc Malay K. Nandy}\footnote{{\tt mknandy@iitg.ac.in} (Corresponding Author)}\\{\normalsize\it Department of Physics, Indian Institute of Technology Guwahati}\\ {\normalsize\it Guwahati 781 039, India}
}

\date{\normalsize (June 7, 2025)}

\maketitle

\begin{abstract}
We explore the inflationary evolution and observational viability of the $\alpha$-attractor E-model in the framework of warm inflation, focusing on both weak and strong dissipative regimes, with a dissipation coefficient linear in temperature. In the strong regime, we account for the growth of inflaton fluctuations due to coupling with the radiation bath via two different forms for the dissipation enhancement function: one associated with plateau-like potentials, and another motivated by the warm little inflation scenario. Employing slow roll conditions, we analytically derive the expressions for the key inflationary observables, the spectral index $n_s$ and the tensor-to-scalar ratio $r$, in both dissipative regimes. The resulting theoretical trajectories on the $n_s$--$r$ plane are then juxtaposed with the contour plots obtained from Planck 2018 data in order to constrain the model parameter. Our analysis shows that the warm $\alpha$-attractor E-model remains compatible with observations in both dissipative regimes, with dissipation playing a crucial role in shifting the predictions and enlarging the viable parameter space, highlighting observational robustness of the model when extended to warm inflation.

\end{abstract}

\maketitle

\tableofcontents

\section{Introduction}
\label{intro}

The inflationary paradigm has become a cornerstone of modern cosmology, offering elegant solutions to the horizon, flatness, and monopole problems inherent in the standard Big Bang model~\cite{guth1981inflationary,Linde-inflation1982}. Traditional models of inflation, often referred to as {\em cold inflation}~\cite{Sato:1980yn,Linde-inflation1982,linde1983chaotic,linde1984inflationary,guth1984inflationary,Albrecht1982A,albrecht1982reheating,HAWKING198235}, posit that the universe underwent a rapid accelerated expansion driven by a scalar field (the inflaton) in a supercooled vacuum state. Perhaps the most fascinating aspect of inflation is that it not only offers a causal explanation for the temperature anisotropies observed in the cosmic microwave background (CMB)~\cite{Larson:2010gs,Bennett:2010jb,Jarosik:2010iu,Hinshaw:2012aka,Planck:2013pxb,Planck:2015xua}, but also provides a natural mechanism for the formation of large-scale structure (LSS) in the Universe. This is because quantum fluctuations during the inflationary epoch can seed the primordial density perturbations responsible for structure formation~\cite{Starobinsky:1979,Mukhanov:1981,Hawking:1982,Guth:1982,Bardeen1983,Starobinsky:1982}.

In the standard cold inflationary scenario, a separate reheating phase is required to reheat the universe and initiate the hot Big Bang phase~\cite{Linde-inflation1982}. Initially, under the slow-roll approximation~\cite{liddle:1994,LiddleLyth2000,KolbTurner1990}, the Universe experiences a period of rapid accelerated expansion, during which its energy density is dominated by a scalar field known as the inflaton. This is followed by the reheating phase~\cite{kofman1994reheating,kofman1996origin,albrecht1982reheating,abbott1982particle,allahverdi2010reheating,amin2015nonperturbative}, wherein the inflaton oscillates around the minimum of its potential, transferring its energy to radiation and thereby initiating the radiation-dominated era of the standard hot big bang cosmology.

An alternative framework, known as {\em warm inflation}, addresses these issues by allowing a continuous transfer of energy from the inflaton field to a radiation bath during the inflationary expansion~\cite{Berera-1995,Berera-2001,Berera_2023}. In warm inflation, dissipative effects modify both the background dynamics and the generation of primordial perturbations. Thermal fluctuations in the radiation bath can dominate over quantum fluctuations, potentially leading to distinct predictions for key cosmological observables~\cite{BERERA2000,Bastero_2016}. Moreover, warm inflation permits steeper potentials and alleviates the $\eta$-problem, making it a compelling candidate for embedding within high-energy physics frameworks, such as supergravity or string-inspired models~\cite{BERERA2000}. 

Recent studies have demonstrated that warm inflation can accommodate observational constraints from Planck data \cite{Kamali_2015,Setare2015,Kamali2016,Visinelli_2016,Motaharfar2016,Panotopoulos2015,Kamali2018,Jawad2017,Jawad-2017,Herrera2015,AlHallak2023,Visinelli_2016}, making it a viable framework for explaining cosmic inflation while simultaneously generating the seeds of cosmic structure. It is also known for rescuing several primordial potentials ruled out in the cold inflation scenario with recent cosmic microwave background data by decreasing the tensor-to-scalar ratio \cite{Benetti2017}.

Among the broad class of inflationary potentials studied in the literature, the $\alpha$-attractor models have received considerable attention due to their robust predictions and geometric origin in supergravity~\cite{Kallosh:2013,Roest:2014}. In particular, the E-model---a subset of $\alpha$-attractors characterized by a non-polynomial potential---has been shown to be compatible with the latest observational data in the cold inflation scenario. However, the implications of embedding the E-model within the warm inflationary framework remain largely unexplored.

In this work, we therfore investigate the dynamics of the E-model in the context of warm inflation. We consider a temperature-dependent dissipative coefficient ($\Gamma/3H$)~\cite{Bastero-Gil_2013,Zhang_2009} of the form $\Gamma \propto T$, motivated by interactions between the inflaton and light fields in supersymmetric setups~\cite{Bastero_2016}. The analysis is carried out in the Einstein frame under the slow-roll approximation, covering both the weak and strong dissipative regimes. We derive the inflationary observables, namely, the scalar spectral index $n_s$, the tensor-to-scalar ratio $r$, and the scalar power spectrum and compare them against the latest constraints from the Planck 2018 data and modified Planck data ~\cite{Planck2018,PRL-Planck}. We also identify the allowed parameter space that yields sufficient e-folds and ensures the consistency of warm E-model inflation with observations.

Our analysis demonstrates that the warm inflationary realization of the E-model is fully compatible with current observational constraints, including those from Planck 2018. These findings significantly extend the phenomenological landscape of warm inflation by establishing the $\alpha$-attractor E-model as a compelling and realistic candidate for describing the early universe. The interplay between dissipative dynamics and the attractor structure enables successful inflation while maintaining agreement with key observables, the spectral index $n_s$ and tensor-to-scalar ratio $r$.

The rest of the paper is organized as follows.  In Section~\ref{alpha}, we provide a brief overview of $\alpha$-attractor models and outline their relevance in inflationary cosmology. Section~\ref{dynamics} introduces the general formalism of warm inflation, which sets the stage for the analysis of the E-model in this context.  In Section~\ref{dissipation-regimes}, we present the implementation of warm inflation within the $\alpha$-attractor E-model framework, considering both weak and strong dissipative regimes, with a dissipation coefficient linear in temperature. Section~\ref{planck} is dedicated to examining the observational viability of the warm $\alpha$-attractor E-model using Planck 2018 data. Finally, Section~\ref{concl} concludes the paper with a discussion.

\section{History of $\alpha$-attractors}
\label{alpha}

The concept of $\alpha$-attractors, introduced by Kallosh, Linde, and Roest~\cite{Kallosh2013}, provides a unifying framework for a broad class of inflationary models within superconformal supergravity. These models are characterized by a parameter $\alpha$, which is inversely proportional to the curvature of the inflaton's Kähler manifold. In the limit of small curvature (large $\alpha$), the predictions for the scalar spectral index $n_s$ and tensor-to-scalar ratio $r$ align with those of chaotic inflation models, such as $V(\phi) \propto \phi^2$. Conversely, for large curvature (small $\alpha$), the models converge to a universal attractor regime characterized by $n_s = 1 - \frac{2}{N}$ and $r=\frac{12\alpha}{N^2}$, where $N$ is the number of e-folds. 
This attractor behavior renders the inflationary predictions remarkably stable across a wide range of potential shapes, aligning well with Planck observational data. The framework provides a unifying geometric foundation for various inflationary models, including the Starobinsky model and E-models, highlighting the natural emergence of plateau-like potentials in such theories.
The versatility of $\alpha$-attractors makes them particularly relevant for warm inflation scenarios, where the interplay between thermal effects and the geometry of the scalar manifold can significantly influence the inflationary dynamics.

Nozari and Rashidi~\cite{Nozari:2018} explored the observational viability of an inflationary scenario based on an E-model potential with nonminimal derivative coupling between the scalar field and gravity. They considered two distinct cases, one with a constant coupling and the other with E-model nonminimal derivative coupling, both with the E-model potential. With these two cases they demonstrated that, in contrast to conventional single-field $\alpha$-attractor models that exhibit an attractor point in the limits of large e-folding number $N$ and small $\alpha$, their setup yielded an attractor line in the $r$-$n_s$ plane under the same limits. By comparing their theoretical predictions with Planck 2015 data, they successfully constrained the model parameters and showed consistency with observations for all values of the mass scale in the model. Furthermore, their analysis revealed that the nonminimal derivative $\alpha$-attractor model allows for a small sound speed, which can result in sizable primordial non-Gaussianities.

Salamate et al.~\cite{Salamate:2019} investigated the observational viability of an inflationary scenario based on the E-model $\alpha$-attractor potential with nonminimal derivative coupling in the context of a braneworld framework. By examining the model predictions for the scalar spectral index $n_s$ and tensor-to-scalar ratio $r$, they confronted theoretical expectations with Planck data to constrain the free parameters of the model. Additionally, they explored the implications for reheating by computing the reheating temperature, thereby establishing a consistent connection between inflation and the subsequent thermal history. Their analysis demonstrated that the model is compatible with current cosmological observations and provides a viable setting for describing early universe inflation with nontrivial gravitational couplings. 

In an influential development, Kallosh and Linde \cite{Kallosh2013b} proposed a superconformal extension of the Starobinsky inflationary model, demonstrating that the celebrated $R + R^2$ theory can be derived from a conformally invariant scalar field framework with spontaneous symmetry breaking. This construction allows for a consistent embedding of inflationary models into supergravity, maintaining the desirable inflationary dynamics while enhancing theoretical stability along the inflationary trajectory. Their work provides a unifying geometric origin for a class of inflationary potentials---including the $\alpha$-attractors and E-models---underscoring the natural emergence of plateau-like potentials within fundamental theories of high-energy physics.

Thus, a significant advancement in inflationary model building was made through the introduction of $\alpha$-attractor models, which unify a broad class of inflationary potentials under a single framework with universal observational predictions. In particular, Kallosh and Linde \cite{Kallosh:2013} demonstrated that models based on conformal and superconformal symmetry exhibit attractor behavior in the $(n_s, r)$ plane, largely independent of the detailed form of the inflaton potential. These models, including the so-called E-models, naturally predict a scalar spectral index $n_s \approx 1 - 2/N$ and a tensor-to-scalar ratio $r \sim 12\alpha/N^2$, where $N$ is the number of e-folds and $\alpha$ encodes the curvature of the Kähler manifold in the supergravity embedding. Their work establishes a theoretical foundation for the robustness of $\alpha$-attractors and motivates their application in extended frameworks, such as warm inflation, where additional dissipative effects are present.

Galante et al.~\cite{Galante:2014} presented a unifying framework for various inflationary models, including $\alpha$-attractors, $\xi$-attractors (such as Higgs inflation), and induced inflation models. They demonstrated that the universal predictions of these models stem from a common feature: a pole of order two in the kinetic term of the inflaton field when expressed in the Einstein frame. This pole structure leads to robust inflationary predictions that are largely independent of the specific form of the potential. They further introduced a class of ``special attractors'' characterized by a specific relation between the parameters $\alpha$ and $\xi$, namely $\alpha = 1 + \frac{1}{6\xi}$. This relation bridges the gap between $\alpha$- and $\xi$-attractors, showing that they can be viewed as different manifestations of a single underlying framework. Notably, in this unified approach, there is no theoretical lower bound on the tensor-to-scalar ratio $r$, allowing for a wide range of inflationary predictions consistent with observational data.

Kallosh and Linde~\cite{Kallosh:2015zsa} explored the robustness of $\alpha$-attractor models in light of observational data from the Planck satellite and potential constraints from the Large Hadron Collider (LHC). These models, rooted in supergravity with logarithmic Kähler potentials, naturally predict inflationary observables that align well with Planck measurements. They extended the framework to incorporate dark energy and supersymmetry breaking, demonstrating that $\alpha$-attractors can accommodate a de Sitter vacuum with spontaneously broken supersymmetry. This unification suggests that the same theoretical structure can describe both the early inflationary universe and its current accelerated expansion, providing a compelling link between cosmology and high energy physics.

Thus, the importance of the $\alpha$-attractor model is significant in modeling the early phase of the universe, as it offers a theoretically consistent and observationally supported framework for inflation. Its ability to yield robust predictions for the scalar spectral index and tensor-to-scalar ratio, which remain stable across a range of model parameters, makes it a particularly attractive candidate for early-universe cosmology. Moreover, the geometric origin of the attractor behavior, stemming from the hyperbolic field space associated with the Kähler potential, enhances its theoretical appeal by linking inflationary dynamics to the underlying structure of supergravity.

Motivated by these strengths, we turn our attention to studying the dynamics of the $\alpha$-attractor E-model within the context of warm inflation. Unlike the standard cold inflation scenario, warm inflation allows for dissipative effects during inflation, leading to a thermal radiation bath coexisting with the inflaton field. This framework can potentially alleviate issues related to reheating and may provide a better fit to the observational data in certain parameter regimes. In this work, we investigate how the warm E-model of inflation behaves under such dynamics and analyze its predictions in light of the latest Planck 2018 observational constraints, particularly focusing on the scalar spectral index $n_s$ and the tensor-to-scalar ratio $r$.

\section{Warm inflation in the $\alpha$-attractor E-model}
\label{dynamics}

In the context of $\alpha$-attractor models of inflation, a canonically normalized scalar field $\phi$ is identified as the inflaton with a potential, 
\begin{equation}
\label{potential}
V(\phi) = \lambda M_{\rm P}^4 \left[1 - e^{-\sqrt{\frac{2}{3\alpha}} \frac{\phi}{M_{\rm P}}} \right]^2,
\end{equation}
known as the E-model.

The energy density $\rho_\phi$ and pressure $P_\phi$ associated with the canonically normalized inflaton field are given by
\begin{equation}
\label{inflaton}
\rho_{\phi} = \frac{1}{2} \dot{\phi}^2 + V(\phi)
\end{equation}
and
\begin{equation}
\label{pressure}
P_{\phi} = \frac{1}{2} \dot{\phi}^2 - V(\phi).
\end{equation}

In the Friedmann-Lema\^itre-Robertson-Walker (FLRW) spacetime, the energy density and pressure of the inflaton field, given in Eqs.~\eqref{inflaton} and \eqref{pressure}, serve as the primary source of the gravitational field during standard (cold) inflation.
The dynamics of the spacetime background is governed by the Friedmann equation,
\begin{equation}
\label{friedmann1}
H^2 = \frac{\rho_\phi}{3 M_{\rm P}^2} = \frac{1}{3 M_{\rm P}^2} \left( \frac{1}{2} \dot{\phi}^2 + V(\phi) \right),
\end{equation}
where $M_P=\frac{1}{\sqrt{8\pi G}}$ is the reduced Planck mass, \(H = \frac{\dot{a}}{a}\) is the Hubble parameter, and \(a(t)\) is the scale factor.

On the other hand, in the warm inflation scenario, the presence of a thermal radiation bath modifies this equation as the inflaton field dissipates its energy into a thermal bath during inflation itself. The total energy density is now given by \(\rho_{\text{tot}} = \rho_\phi + \rho_r\),  and the Friedmann equation becomes
\begin{equation}
\label{fried1}
H^2 = \frac{1}{3 M_{\rm P}^2} \left( \rho_\phi + \rho_r \right),
\end{equation}
with $\rho_r$ the radiation energy density.

Since the warm inflation framework allows inflation to proceed in the presence of a thermal bath, a dissipative coefficient \(\Gamma(\phi, T)\) is introduced to account for this interaction, so that the evolutions of the inflaton energy density $\rho_{\phi}$ and the radiation energy density $\rho_r$ can be modelled as~\cite{MOSS1985,Berera-1995,Berera_1996}
\begin{equation}\label{KG}
\dot \rho_{\phi}+3H(\rho_{\phi}+P_{\phi})=-\Gamma\dot\phi^2
\end{equation}
and
\begin{equation}
\label{reheat}
\dot\rho_r+3H(\rho_r+p_r)=\Gamma\dot\phi^2,
\end{equation}
indicating that radiation is continually produced by the dissipative dynamics of the inflaton, with the radiation equation of state, $p_r=\frac{1}{3}\rho_r$.

The basic premise of warm inflation being the presence of a thermalised radiation in the inflationary period, its energy density given by
\begin{equation}\label{radiation}
\rho_r=C_* T^4,
\end{equation}
where $T$ is the temperature and $C_*=\frac{\pi^2}{30}g_*$, with $g_*= 228.75$ \cite{Panotopoulos2015} the effective number of degrees of freedom.

Using \ref{inflaton}, equation \ref{KG} gives the dynamics of the inflaton field as
\begin{equation}
\label{KG1}
\ddot\phi+3H\dot\phi+V'(\phi)=-\Gamma\dot\phi,
\end{equation}
which, upon neglecting $\ddot\phi$ in the slow roll approximation, gives
\begin{equation}
\label{KG1a}
3H\dot\phi+V'(\phi)=-\Gamma\dot\phi.
\end{equation}

In the warm inflationary scenario, the reduction in radiation energy density due to the Hubble expansion is assumed to be continuously replenished by radiation produced through the decay of the inflaton field \cite{Berera_Fluc_1995}. As a result, the term $\dot\rho_r$ can be neglected, assuming $\dot\rho_r \ll 4H\rho_r$ and $\dot\rho_r \ll \Gamma\dot\phi^2$~~\cite{MOSS1985,Berera_1996,Berera:1996fm,Berera_Fluc_1995,Berera:1999ws,Hall_2004}. Thus, equation~~\ref{reheat} approximates to
\begin{equation}
\label{reheat2}
4H\rho_r=\Gamma\dot\phi^2.
\end{equation}

Eliminating $\dot\phi$ between equations \ref{KG1} and \ref{reheat2}, and substituting for $\rho_r$ in equation \ref{radiation}, we obtain the radiation temperature as
\begin{equation}
\label{temperature}
T=\left(\frac{\Gamma V'^2}{36C_*H^3(1+Q)^2}\right)^{1/4},
\end{equation}
where $Q$ is the dissipation ratio, defined by
\begin{equation}
\label{R}
Q = \frac{\Gamma}{3H}.
\end{equation}

Various slow roll parameters in warm inflation may be defined as \cite{Moss_2008,Hall_2004}
\begin{align}
\label{slow-roll}
\epsilon=\frac{M_P^2}{2}\left(\frac{V^\prime}{V}\right)^2, \qquad \eta=M_P^2\left(\frac{V''}{V}\right), \qquad
\beta=M_P^2\left(\frac{\Gamma^\prime V^\prime}{\Gamma V}\right),
\end{align}
with the slow roll conditions
\begin{align}
\label{condition}
\epsilon &\ll 1+Q, \qquad \eta \ll 1+Q, \qquad 
\beta \ll 1+Q.
\end{align}

The key physical quantities in this study are assessed at the epoch when the relevant cosmic microwave background (CMB) modes exit the Hubble horizon, that typically occurs about 50-60 e-folds before the end of inflation. At this horizon-crossing, the inflaton fluctuation spectrum is characterized by a phase space occupation number $n_*$. Its value depends critically on the interaction of the inflaton with the surrounding thermal bath. In scenarios where this interaction is negligible, the fluctuations remain in the vacuum state, corresponding to the Bunch-Davies limit $(n_* = 0)$. On the other extreme, strong thermal contact with the radiation bath drives the fluctuations toward a Bose-Einstein distribution at temperature $T_*$, leading to $n_* \approx \frac{1}{e^{H_*/T_*} - 1}$. Our analysis focuses on this thermally populated regime.

Consequently, the amplitude of power spectrum can be written as~\cite{bastero2018,Graham:2009,Bastero-Gil:2009-model,Hall_2004,Ramos:2013,Taylor:2000,deOliveira:2001,Visinelli_2016}
\begin{equation}\label{Power}
\Delta\mathcal R =\left(\frac{{H_*}^2}{2 \pi \dot \phi_*}\right)^2 \left[1+ 2n_* +\left(\frac{T_*}{H_*}\right)\frac{2\sqrt{3}\pi Q_*}{\sqrt{3 + 4\pi Q_*}}\right]G(Q_*),
\end{equation}
where the enhancement function $G(Q_*)$ quantifies the growth of inflaton fluctuations due to their coupling with the radiation bath.  This enhancement stems from the temperature dependence of the dissipation coefficient, which governs energy transfer between the inflaton and the thermal bath. The precise form of $G(Q_*)$ is not analytically tractable in general and must be obtained via numerical methods. Following the approach outlined in \cite{Bastero_2016,bastero2018}, we shall incorporate $G(Q_*)$ in the context of the $\alpha$-attactor E-model potential considered in this study.

\section{Dissipative dynamics of warm $\alpha$-attractor E-model}
\label{dissipation-regimes}

To explore the weak and strong regimes of warm inflation in the context of the $\alpha$-attactor E-model, we model the dissipation coefficient as
\begin{equation}
\label{dissipation}
\Gamma=C_1 T,
\end{equation}
where $C_1$ is a constant parameter.

\subsection{Weak dissipation}
\label{weak-reg}

In the weak dissipative regime, characterized by $\Gamma \ll 3H$, the influence of the dissipative coefficient is negligible compared to the Hubble expansion rate. With this approximation, equation \ref{temperature} reduces to
\begin{align}
\label{T-weak}
T =\left(\frac{C_1 V^{\prime 2}}{36C_* H^3}\right)^{1/3}
  =\left(\frac{2\sqrt 3 C_1}{9 \alpha C_* M_P^2}\frac{e^{-2x}}{1-e^{-x}}\right)^{1/3},
\end{align}
upon substituting from equation \ref{potential}, with $x=\sqrt{\frac{2}{3\alpha}\frac{\phi}{M_P}}$.

Assuming the slow-roll condition, $\frac{1}{2}\dot{\phi}^2 \ll V(\phi)$, and that the energy density of radiation is subdominant compared to that of the inflaton, i.e., $\rho_r \ll \rho_\phi$, equations \ref{fried1} and \ref{inflaton} simplify, yielding
\begin{equation}
\label{fried1a}
H=\left(\frac{V}{3M_P^2}\right)^{1/2}.
\end{equation}

In this regime, neglecting the dissipation term by assuming $\Gamma \ll 3H$, equation \ref{KG1a} reduces to
\begin{equation}
\label{phidot}
\dot{\phi} = -\frac{V'(\phi)}{3H} = -\frac{1}{M_P} \frac{2\sqrt{\frac{2}{3\alpha} e^{-x} (1- e^{-x})}}{3H}.
\end{equation}

By applying the $\alpha$-attractor E-model potential given in equation \ref{potential}, the slow-roll parameters, as defined in equation \ref{slow-roll}, take the forms
\begin{equation}
\label{epsilon}
\epsilon = \frac{4}{3 \alpha} \frac{1}{(e^{-x}-1)^2},
\end{equation}
\begin{equation}
\label{eta}
\eta = \frac{4}{3 \alpha} \frac{2-e^x}{(e^{-x}-1)^2},
\end{equation}
and
\begin{equation}
\label{beta-weak}
\beta = \frac{4}{9 \alpha} \frac{(1-2e^x)}{(e^{x}-1)^2}.
\end{equation}

Given that $Q \ll 1$ in the weak dissipation regime, inflation is considered to end when the slow-roll parameter satisfies $\epsilon = 1$. Thus equation \ref{epsilon} gives
\begin{align}
 x_e   &= \ln\left(\frac{2}{\sqrt{3\alpha}} +1\right),
 \end{align}
or, equivalently
\begin{align}
 \phi_e &= \sqrt{\frac{3\alpha}{2}}\ln\left(\frac{2}{\sqrt{3\alpha}} +1\right)\ M_P
\end{align}
at the end of inflation (denoted by the subscript $_e$).

The number of e-folds can be calculated as
\begin{align}
\label{ef}
N = \frac{1}{M_P^2} \int_{\phi_e} ^{\phi_*} \frac{V}{V'} d\phi
  = \frac{3\alpha}{4}\int_{x_e} ^{x_*} (e^x -1) dx.
\end{align}
Since we expect $\phi_*\gg\phi_e$ (equivalently, $x_*\gg x_e$), equation \ref{ef} gives
\begin{equation}
\label{xstar}
e^{x_*}= \frac{4N}{3\alpha},
\end{equation}
that implies
\begin{equation}
\label{phi_*-weak}
\phi_*= \sqrt{\frac{3\alpha}{2}} \ln \left(\frac{4N}{3\alpha}\right)\ M_P,
\end{equation}
where the subscript $_*$ denotes the value at Hubble crossing.

\begin{figure*}
\includegraphics[scale=0.4]{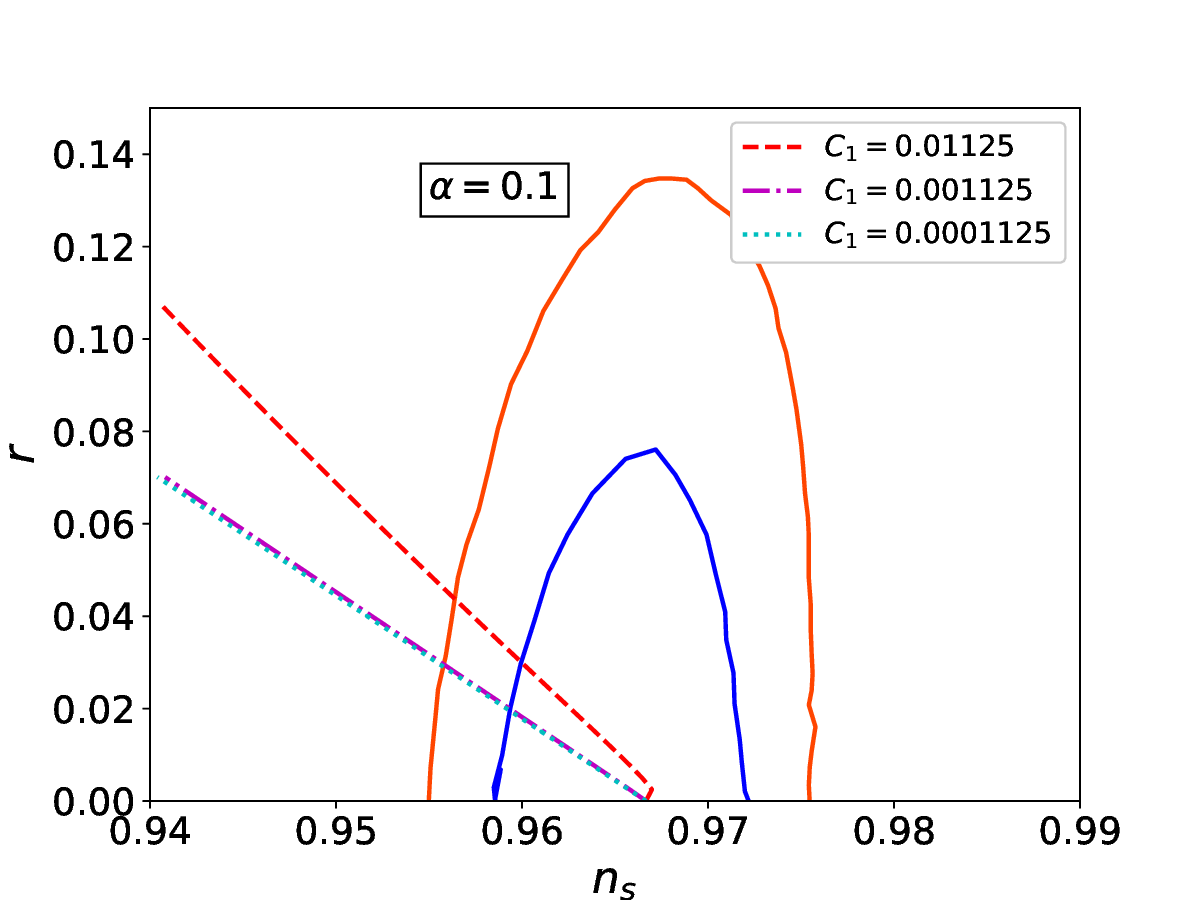}
\includegraphics[scale=0.4]{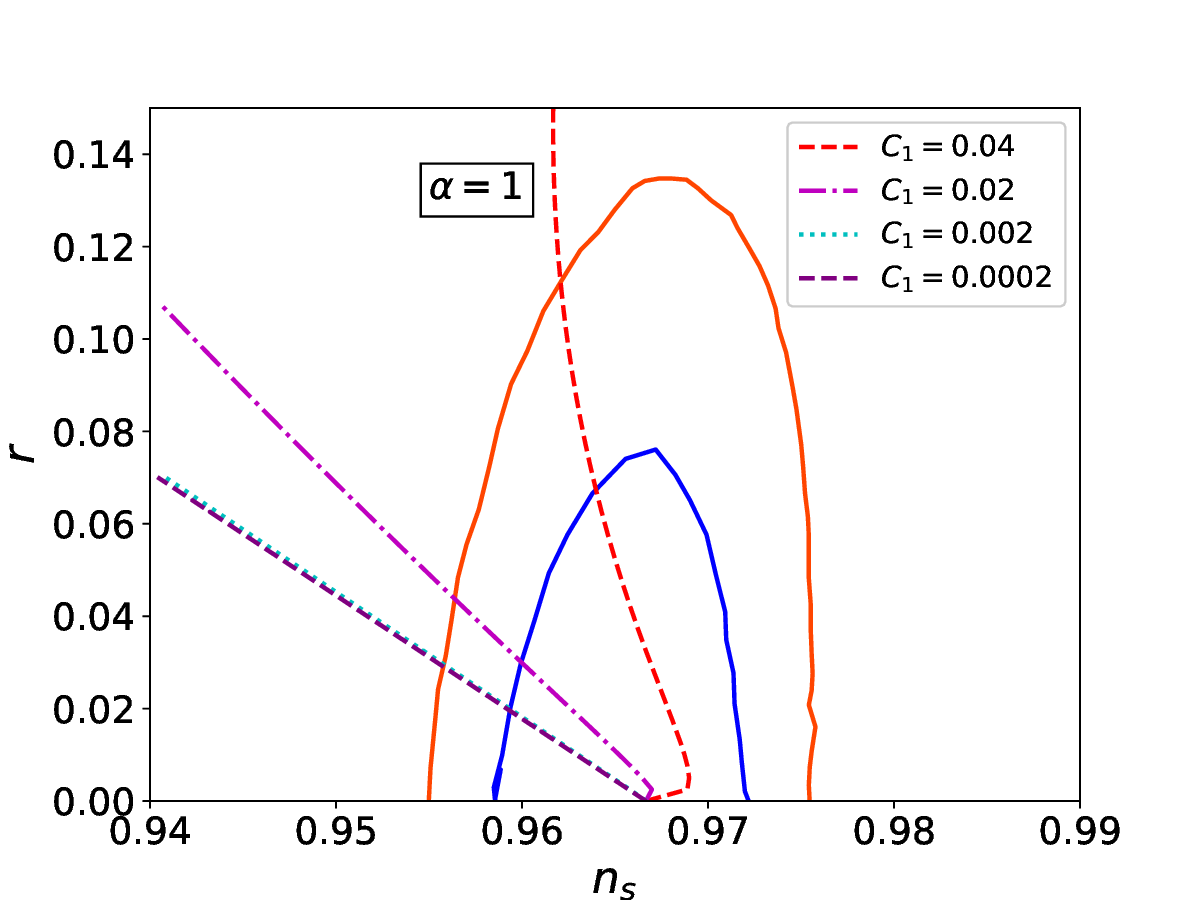}
\includegraphics[scale=0.4]{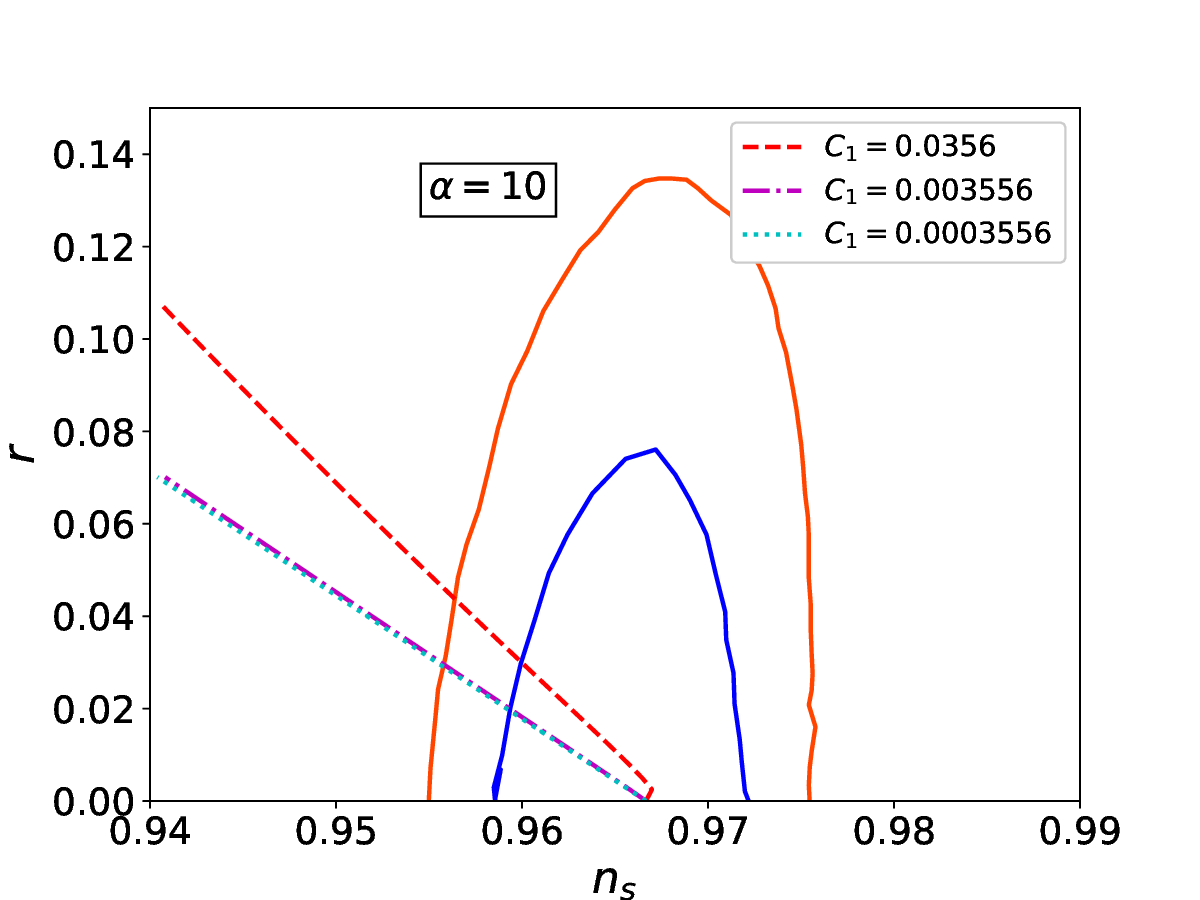}
\includegraphics[scale=0.4]{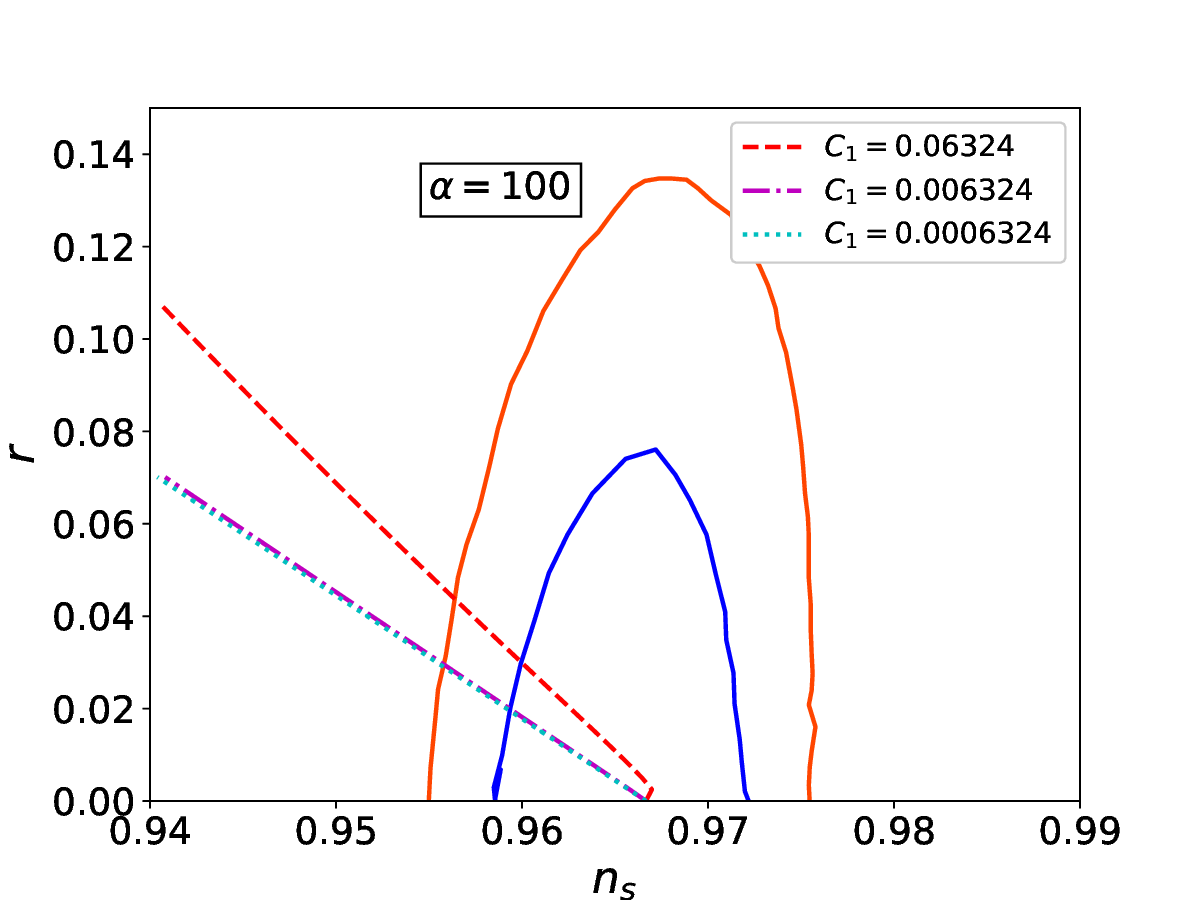}
\caption{\small Trajectories in the $n_s$--$r$ plane with different values of $\alpha$ (in the admissible range $10^{-2} < \alpha < 10^4$) in the weak regime of $\alpha$-attractor E-model overlaid on the observational constraints provided by the Planck 2018 data.}
\label{fig-1}
\end{figure*}

Making use of the relation \ref{xstar}, we can obtain the slow roll parameters given by equations \ref{epsilon}, \ref{eta}, and \ref{beta-weak} in terms of number of e-folds $N$ as
\begin{equation}
\label{enb}
\epsilon = \frac{3 \alpha}{4N^2},\ \ \ \
\eta = - \frac{1}{N},\ \ \ \
\beta = -\frac{2}{3N}.
\end{equation}

Futhermore, using equations \ref{R}, \ref{dissipation}, and \ref{T-weak}, the expression for $Q$ in weak regime turns out to be
\begin{equation}
\label{Q-weak}
Q = \frac{1}{3} \left(\frac{2{C_1}^4}{C_*}\frac{1}{\alpha \lambda}\frac{1}{(e^{x}-1)^2}\right)^{1/3}
\end{equation}

Now using equation \ref{xstar} in \ref{Q-weak}, we obtain $Q$ at horizon exit as
\begin{equation}
\label{Qstar}
Q_* =  \left(\frac{{C_1}^4}{3C_*}\frac{1}{\lambda}\frac{\alpha}{8N^2}\right)^{1/3}.
\end{equation}

We further have the COBE normalisation condition \cite{Bezrukov_2009}
\begin{equation}
\label{COBE}
\left.\frac{V}{\epsilon}\right|_{\phi=\phi_*}= (0.0276 M_P)^4.
\end{equation}

Using \ref{slow-roll} and \ref{epsilon}, the ratio $\frac{V}{\epsilon}$ can be calculated as
\begin{align}
\label{ratio}
\frac{V}{\epsilon}= \frac{2V^3}{M_P^2 V'^2}= \frac{3 \alpha}{4} \lambda \, M_P^{4} \, e^{2x}(1- e^{-x})^4.
\end{align}

Using \ref{ratio} and \ref{xstar} in \ref{COBE}, and noting that $x_*\gg 1$, we obtain the coupling parameter as
\begin{align}
\label{lambda}
\lambda = (0.0276)^4 \frac{3 \alpha}{4N^2}.
\end{align}

In weak regime, with $Q\ll 1$, the tensor-to-scalar ratio has the form $r=16\epsilon$~\cite{Hall_2004,Ramos:2013}. This expression, upon substituting from equation~\ref{epsilon}, becomes
	\begin{equation}\label{r-N}
		r= \frac{12 \alpha}{N^2}.
	\end{equation}

The expression for the spectral index $n_s$, up to the first order correction of $Q$, is given by \cite{Hall_2004,Ramos:2013}
\begin{align}
n_s = 1-6\epsilon +2 \eta +Q(8 \epsilon -2\beta)
\end{align}
Substituting from \ref{enb}, this expression goes over to
\begin{align}
n_s = 1- \frac{9\alpha}{2 N^2} -\frac{2}{N} +Q\left(\frac{6\alpha}{N^2} + \frac{4}{3N}\right) 
\end{align}
which, upon using \ref{Qstar}, yields
\begin{align}
n_s = 1- \frac{9\alpha}{2 N^2} -\frac{2}{N} + \left(\frac{{C_1}^4}{3C_*}\frac{1}{\lambda}\frac{\alpha}{8N^2}\right)^{1/3} \left(\frac{6\alpha}{N^2} + \frac{4}{3N}\right).
\end{align}
Substituting $\alpha = \frac{1}{12} r N^2$ \ref{r-N}, this expression reduces to
\begin{equation}
\label{ns-weak}
n_s= 1 -\frac{2}{N} -\frac{3r}{8}+\left[\frac{r}{2} +  \left(\frac{{C_1}^4}{288 C_*}\frac{1}{\lambda}r\right)^{1/3}\frac{4}{3N}\right],
\end{equation}
where $\lambda$ can be expressed in terms of $\alpha$ using equation \ref{lambda}.

\subsection{Strong dissipation}
\label{strong-reg}

In the strong dissipation regime, the interaction between the inflaton field and the surrounding thermal bath is sufficiently strong that energy transfer from the inflaton occurs at a rate much faster than the Hubble expansion, that is, $\Gamma \gg 3H$, or equivalently, $Q \gg 1$.

Using \ref{dissipation}, equation \ref{temperature} therefore modifies to
\begin{equation}\label{T-strong}
T=\left(\frac{1}{4C_*C_1}\right)^{1/5}\left(\frac{V'^2}{H}\right)^{1/5},
\end{equation}
and equation \ref{KG1a} leads to
\begin{equation}
\label{KG2}
\dot\phi=-\frac{V'(\phi)}{\Gamma}=-\frac{1}{M_P}\frac{2\sqrt{\frac{2}{3\alpha} e^{-x} (1- e^{-x})}}{\Gamma}
\end{equation}
for the quartic potential \ref{potential}.

By applying the $\alpha$-attractor E-model potential \ref{potential}, the slow-roll parameters \ref{slow-roll} are obtained in the strong regime as
\begin{equation}
\label{epsilon-str}
\epsilon = \frac{4}{3 \alpha} \frac{1}{(e^{-x}-1)^2},
\end{equation}
\begin{equation}
\label{eta-str}
\eta = \frac{4}{3 \alpha} \frac{2-e^x}{(e^{-x}-1)^2},
\end{equation}
and
\begin{equation}
\label{beta-str}
\beta = \frac{1}{5}(2\eta-\epsilon)= \frac{4}{15 \alpha} \frac{(3-2e^x)}{(e^{x}-1)^2}.
\end{equation}
Thus, whearas the expressions for $\epsilon$ and $\eta$ remain the same as in the weak regime,
the expression for $\beta$ differ because of differences in the expressions for the temperature, given by \ref{T-weak} and \ref{T-strong}, in the two regimes.

Using equations \ref{R}, \ref{dissipation}, and \ref{T-strong}, the expression for $Q$ in the strong regime turns out to be
\begin{align}
\label{Q-strong}
Q=\left[\left(\frac{2}{3}\right)^3 \frac{{C_1}^4}{4 C_* \lambda \alpha}\frac{e^{-2x}}{(1-e^{-x})^4}\right]^{1/5}
\end{align}

At the end of inflation $\epsilon_{e}=Q$, thus we get from \ref{Q-strong} and \ref{epsilon-str}
\begin{align}
\label{phi-end-strong}
\phi_e = \sqrt{\frac{3\alpha}{2}}\frac{1}{8}\ln \left[\frac{2^7}{3^2\alpha^5} \frac{C_* \lambda}{C_1^4}\right]M_P
\end{align}

In the large field limit, $x_*\gg 1$, equation \ref{Q-strong} yields
\begin{align}
\label{Q-star}
Q_*= \left[\left(\frac{2}{3}\right)^3 \frac{{C_1}^4}{4 C_* \lambda \alpha}\frac{1}{e^{2x_*}}\right]^{1/5}
\end{align}

The number of e-folds in this regime, given by
\begin{align}
N = \int _{\phi_e}^{\phi_*} \frac{V}{V'}Q d\phi 
  = \sqrt{\frac{3\alpha}{8}} \sqrt{\frac{3\alpha}{2}} \left(\frac{2}{3}\right)^{3/5} \left(\frac{C_1^4}{4 C_* \lambda}\right)^{1/5} \int _{x_e}^{x_*}e^{3x/5} dx,
\end{align}
is dominated by the large field region, so that it approximates to
\begin{align}
N=\frac{5 \alpha}{4} \left(\frac{2}{3}\right)^{3/5} \left(\frac{C_1^4}{4 C_* \lambda}\right)^{1/5}e^{3x_*/5}.
\label{N-strong}
\end{align}
This expression yields the value of the inflaton field,
\begin{align}
\phi_*= \frac{1}{3}\sqrt{\frac{3\alpha}{2}} \ln\left[\left(\frac{N}{5 \alpha}\right)^5 2^7 3^3\left(\frac{4 C_* \lambda}{C_1^4}\right)\right] M_P,
\end{align}
at Hubble crossing.

Now, we can write the slow roll parameters \ref{epsilon-str}, \ref{eta-str} in terms of $N$, the number of e-folds \ref{N-strong}, as
\begin{equation}
\label{slow-strong}
\epsilon = \frac{2^4}{3^3\alpha} \left(\frac{5\alpha}{4N}\right)^{10/N}\left(\frac{C_1^4}{4C_*\lambda N}\right)^{2/3}
\end{equation}
and
\begin{align}
\label{slow-strong2}
\eta = -\sqrt{\frac{4\epsilon}{3\alpha}} 
     = -\frac{2^3}{3^2 \alpha} \left(\frac{5\alpha}{4N}\right)^{5/N}\left(\frac{C_1^4}{4C_*N}\right)^{1/3}.
\end{align}

\begin{figure*}
	\includegraphics[scale=0.4]{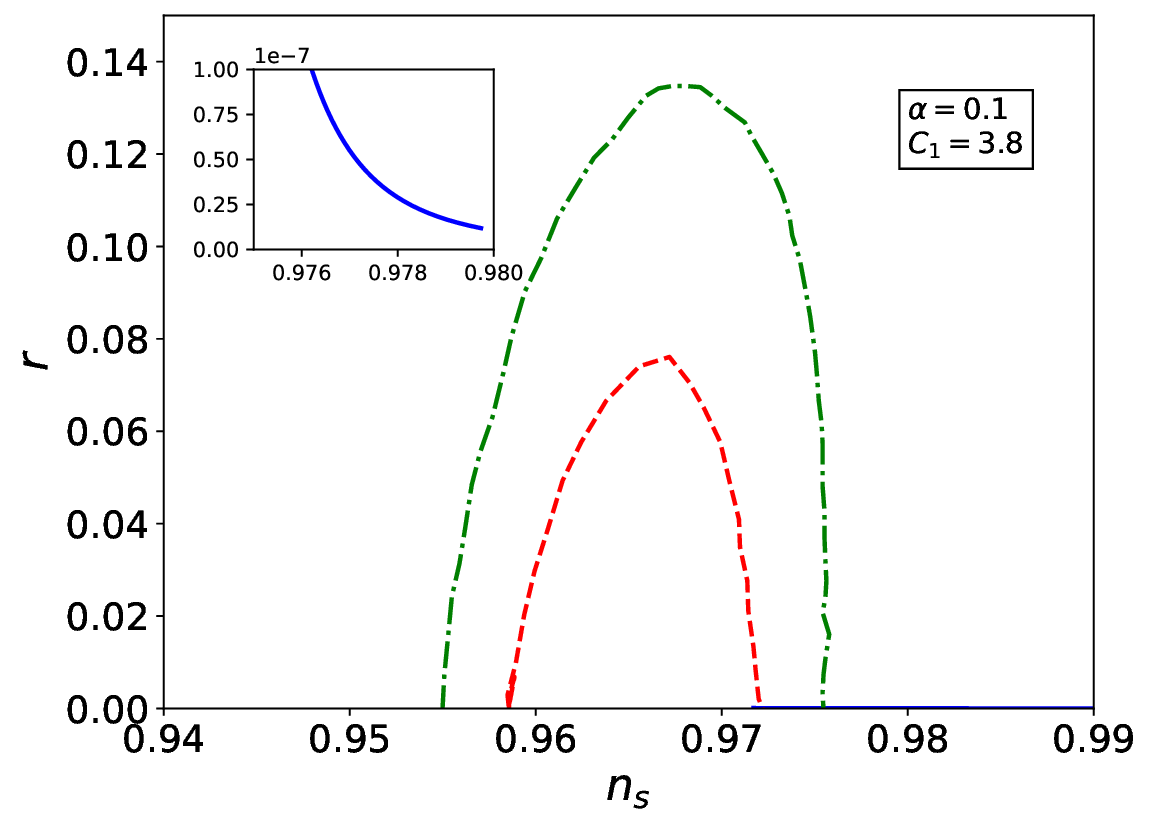}
	\includegraphics[scale=0.4]{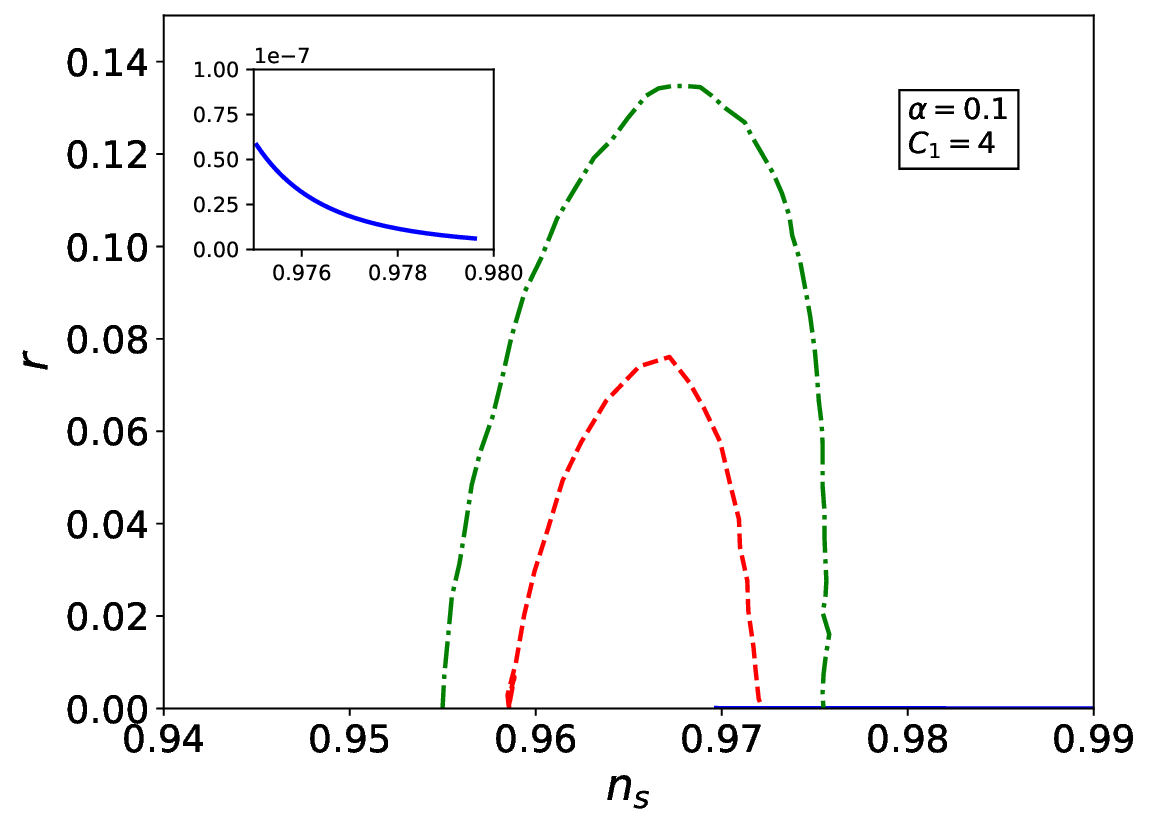}
	\includegraphics[scale=0.4]{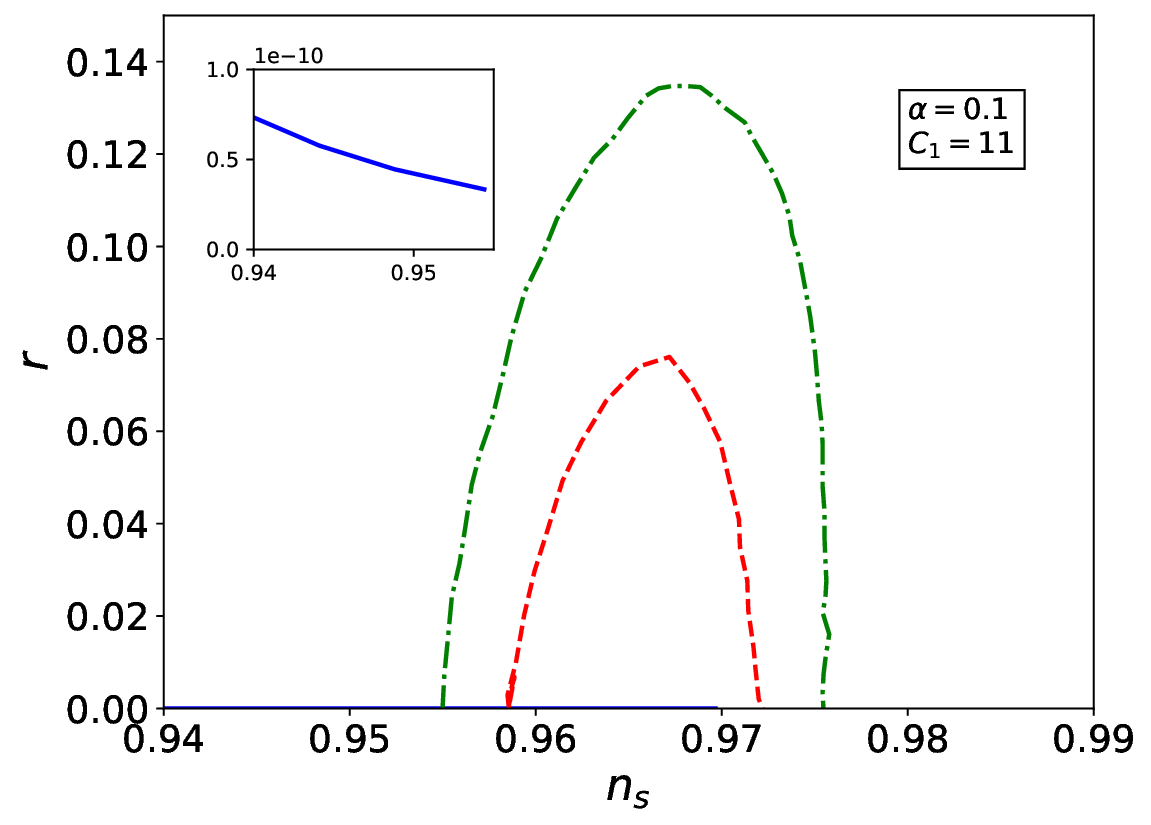}
	\includegraphics[scale=0.4]{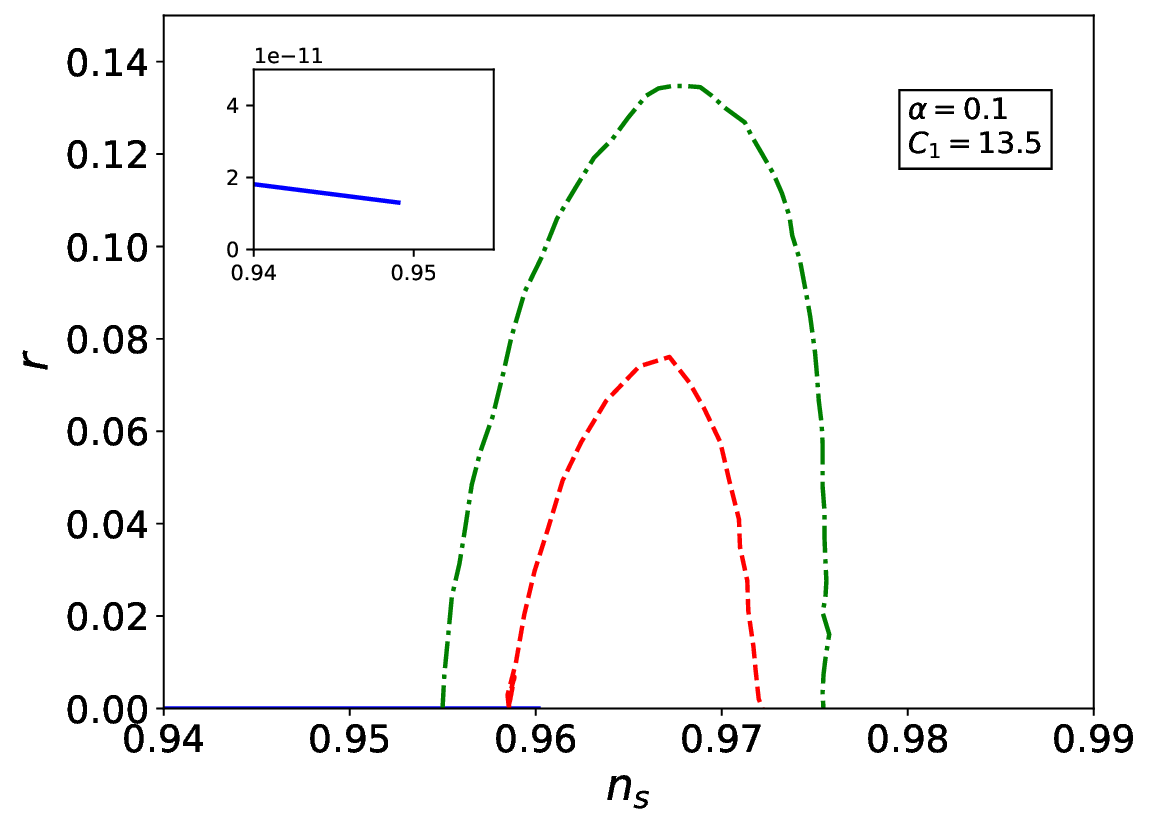}
	\caption{\small Inflationary trajectories in the $n_s$--$r$ plane with different values of $C_1$ for fixed $\alpha=0.1$ (in the admissible range $10^{-2} < \alpha < 10^4$) in the strong regime of $\alpha$-attractor E-model employing the enhancement function $G_1(Q_*)$ for plateau-like potentials: Comparison with Planck 2018 data.}
	\label{fig-2}
\end{figure*}

In the amplitude of power spectrum \ref{Power} for the strong regime, we shall use
\begin{align}
\label{plateau}
	G_1(Q_*)  &\simeq 1 + 0.18 {Q_*}^{1.4} + 0.01 {Q_*}^{1.8}
\end{align}
and
\begin{align}
\label{little}
	G_2 (Q_*) &\simeq 1 + 0.335 {Q_*}^ {1.364} + 0.0185 {Q_*}^ {2.315}
\end{align}
for plateau-like potential \cite{bastero2018} and for warm little inflation \cite{Bastero_2016}, respectively.

Following from \cite{bastero2018}, the spectral index can be written as 
\begin{equation}
n_s -1 = \left.\frac{d \ln \Delta R}{d \ln k}\right|_{k=k_*}= \frac{d \ln \Delta R}{dN}
\end{equation}
where $\ln k= aH =N$.

Using the Friedmann equation \ref{fried1a} and further substituting from \ref{KG1a} and \ref{slow-roll},  the amplitude of power spectrum \ref{Power} can be expressed as
\begin{align}
\Delta R =  \left(\frac{V(\phi_*)(1+ Q_*)^2}{24 M_P^4 \pi^2 \epsilon}\right)\left[1+ 2n_* + \left(\frac{T_*}{H_*}\right)\frac{2\sqrt3 \pi Q_*}{\sqrt{3+4\pi Q_*}}\right]G(Q_*).
\end{align}
For thermalized fluctuations, $n_*\simeq\frac{T_*}{H_*}\gtrsim 1$ and $\frac{T_*}{H_*}= 3\frac{Q_*}{C_1}$ \cite{bastero2018}, so that $1+2n_* = 2\left(\frac{T_*}{H_*}\right)= 6\left(\frac{Q_*}{C_1} \right)$, yielding 
\begin{align} 
\Delta R =  \left(\frac{V(\phi_*)(1+ Q_*)^2}{4 M_P^4 \pi^2 \epsilon}\right)\left(\frac{Q_*}{C_1}\right)\left[1+  \frac{\sqrt3 \pi Q_*}{\sqrt{3+4\pi Q_*}}\right]G(Q_*).
\end{align}
Now, using $\frac{\rho_r}{V}=\frac{\epsilon Q}{2(1+Q)^2}$ that follows from \ref{KG1a}, \ref{reheat2}, and \ref{slow-roll}, the above expression reduces to
\begin{align} 
	\label{del-R}
\Delta R = \frac{5C_1^3}{12 g_* \pi^4 Q_*^2} \left[1+  \frac{\sqrt3 \pi Q_*}{\sqrt{3+4\pi Q_*}}\right]G(Q_*).
\end{align}
Differentiating this expression \ref{del-R} with respect to $Q_*$, we have
\begin{align} \label{ddel-R}
\frac{d \Delta R}{dQ_*} = \frac{5C_1^3}{12 g_* \pi^4 } \left[\frac{G(Q_*)}{Q_*^2}\left(-\frac{2}{Q_*} -\frac{\sqrt 3 \pi }{\sqrt{3+4\pi Q_*}}+ \frac{2\sqrt3 \pi^2 Q_* }{\sqrt{3+4\pi Q_*}}\right)+\frac{1}{Q_*^2}\left(1+\frac{\sqrt3 \pi Q_*}{\sqrt{3+4\pi Q_*}}\right)G'(Q_*)\right]
\end{align}
The spectral index of power spectrum for growing modes is given by \cite{bastero2018,BASTERO-2009,Benetti2017}
\begin{align}
	\label{ns-strong}
n_ s &= 1+ \frac{Q_*}{3+5Q_*} (6\epsilon -2 \eta) \frac{1}{\Delta R} \frac{d \Delta R}{dQ_*}.
\end{align}
Thus, substituting from equations \ref{slow-strong} and \ref{slow-strong2} into \ref{ns-strong}, we obtain
\begin{align}
	n_s = 1+ \frac{Q_*}{3+5Q_*} \left[\frac{2^4}{3^2 \alpha}\left(\frac{5\alpha}{4N}\right)^{5/N}\left(\frac{B}{N}\right)^{1/3}\left\{2\left(\frac{5\alpha}{4N}\right)^2\left(\frac{B}{N}\right)^2-1\right\}\right]\frac{1}{\Delta R}\frac{d \Delta R}{dQ_*},
\end{align}
which, upon substituting from \ref{ddel-R}, yields
\begin{align}
     n_s&= 1+ \frac{Q_*}{3+5Q_*} \left[\frac{2^4}{3^2 \alpha}\left(\frac{5\alpha}{4N}\right)^{5/N}\left(\frac{B}{N}\right)^{1/3}\left\{2\left(\frac{5\alpha}{4N}\right)^2\left(\frac{B}{N}\right)^2-1\right\}\right]\\ \nonumber
     &\frac{1}{\Delta R}\left[\frac{5C_1^3}{12 g_* \pi^4 } \left[\frac{G(Q_*)}{Q_*^2}\left(-\frac{2}{Q_*} -\frac{\sqrt 3 \pi }{\sqrt{3+4\pi Q_*}}+ \frac{2\sqrt3 \pi^2 Q_* }{\sqrt{3+4\pi Q_*}}\right)+\frac{1}{Q_*^2}\left(1+\frac{\sqrt3 \pi Q_*}{\sqrt{3+4\pi Q_*}}\right)G'(Q_*)\right]\right],
\end{align}
that can be transformed to the expression
\begin{align}
\label{ns-final}
     n_s&= 1+ \frac{Q_*}{3+5Q_*} \left[\frac{2^4}{3^2 \alpha}\left(\frac{5\alpha}{4N}\right)^{5/N}\left(\frac{B}{N}\right)^{1/3}\left\{2\left(\frac{5\alpha}{4N}\right)^2\left(\frac{B}{N}\right)^2-1\right\}\right]\\ \nonumber
     &\left[\left(-\frac{2}{Q_*} -\frac{\sqrt 3 \pi }{\sqrt{3+4\pi Q_*}}+ \frac{2\sqrt3 \pi^2 Q_* }{\sqrt{3+4\pi Q_*}}\right)\left(1 + \frac{\sqrt 3 \pi Q_*}{\sqrt{3+4\pi Q_*}}\right)^{-1} + \frac{G'(Q_*)}{G(Q_*)}\right].
\end{align}
with further substitution from \ref{del-R}.

Furthermore, following \cite{bastero2018}, the  tensor-to-scalar ratio is given by 
\begin{equation}
\label{r}
r=\frac{\Delta T}{\Delta R}, \quad \text{where} \quad \Delta T= \frac{2H^2}{\pi^2 M_P^2},
\end{equation}
Using the Friedmann equation \ref{fried1a}, this equation leads to
\begin{align}
\label{r-strong}
  r = \frac{2V}{3\pi^2 M_P^4 \Delta R},
\end{align}
that, upon substituting from \ref{del-R}, finally yields
  \begin{align}
  \label{r-final}
  r= \frac{8Vg_* \pi^2 Q_*^2}{5 C_1^3 M_P^4}\left[\left(1 + \frac{\sqrt 3 \pi Q_*}{\sqrt{3+4\pi Q_*}}\right)^{-1}{G(Q_*)}^{-1}\right]
\end{align}

\begin{figure*}
\centering
\includegraphics[scale=0.5]{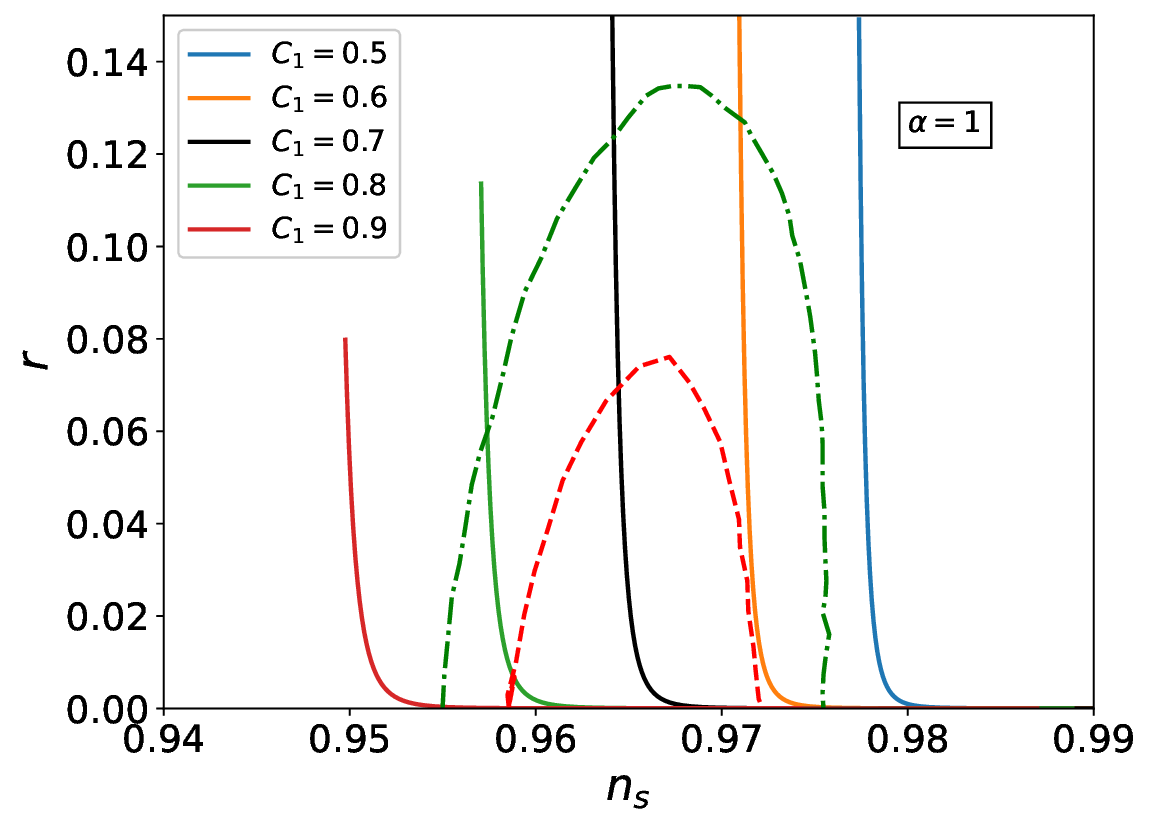}
\caption{\small Trajectories in the $n_s$--$r$ plane with different values of $C_1$ for fixed $\alpha=1$ in the strong regime of $\alpha$-attractor E-model employing the enhancement function $G_1(Q_*)$ for plateau-like potentials: Comparison with Planck 2018 data.}
\label{fig-3}
\end{figure*}

\section{Viability of warm $\alpha$-attractor E-Model by Planck 2018}
\label{planck}

In this section, we investigate the viability of the $\alpha$-attractor E-model in the weak and strong dissipative regimes of warm inflation. In this model, Planck 2018 observational bounds restrict the parameter $\alpha$ in the range $10^{-2} < \alpha < 10^4$ \cite{Planck2018}.

In the weak regime, we use the analytical relation \ref{ns-weak} to obtain theoretical predictions for for the scalar spectral index $n_s$ and tensor-to-scalar ratio $r$. On the other hand, in the strong regime, we employ the two distinct enhancement functions---one corresponding to plateau-like potentials and the other to the warm little inflation scenario, given by \ref{plateau} and \ref{little}, respectively---and obtain theoretical predictions for $n_s$ and $r$ from the relations \ref{ns-final} and \ref{r-final}. These predictions are then rigorously compared against the latest observational constraints from the Planck 2018 data. This analysis allows us to delineate the parameter space where the model remains consistent with current cosmological observations, thereby assessing its feasibility as a candidate for the warm inflationary scenario.

\subsection{Weak dissipative regime}

Using equation~\ref{ns-weak}, we compute and plot the scalar spectral index $n_s$ against the tensor-to-scalar ratio $r$ for a range of values of the parameters $\alpha$ and $C_1$. These trajectories are then overlaid on the observational constraints in the $n_s$--$r$ plane provided by the Planck 2018 data. The comparison allows us to extract viable regions of parameter space consistent with observations. Specifically, the Planck 2018 results impose a stringent upper bound on the combination $\frac{C_1^4}{\alpha}$, which must satisfy $\frac{C_1^4}{\alpha} \leq 1.6 \times 10^{-7}$ to remain within the 68\% and 95\% confidence contours. This constraint enables us to determine an allowed range for the parameter $C_1$ for each fixed value of $\alpha$, ensuring compatibility with current cosmological measurements \cite{Planck2018,aghanim2020planck}.

In Fig.~~\ref{fig-1}, we illustrate the computed values of $r$ versus $n_s$ derived from equation~~\ref{ns-weak}, superimposed with the Planck 2018 observational contours. These plots serve as a diagnostic tool to constrain the model parameter $C_1$ across a representative set of values for $\alpha$, specifically $\alpha = 0.1, 1, 10, 100$. For each case, the corresponding range of $C_1$ values that yield predictions within the Planck bounds is identified. These results are systematically summarized in Table~\ref{tab-1}, which presents the upper bounds on $C_1$ for each chosen value of $\alpha$, highlighting the dependence of the allowed parameter space on the underlying geometry of the scalar field manifold encoded in $\alpha$.

\begin{figure*}
\centering
	\includegraphics[scale=0.5]{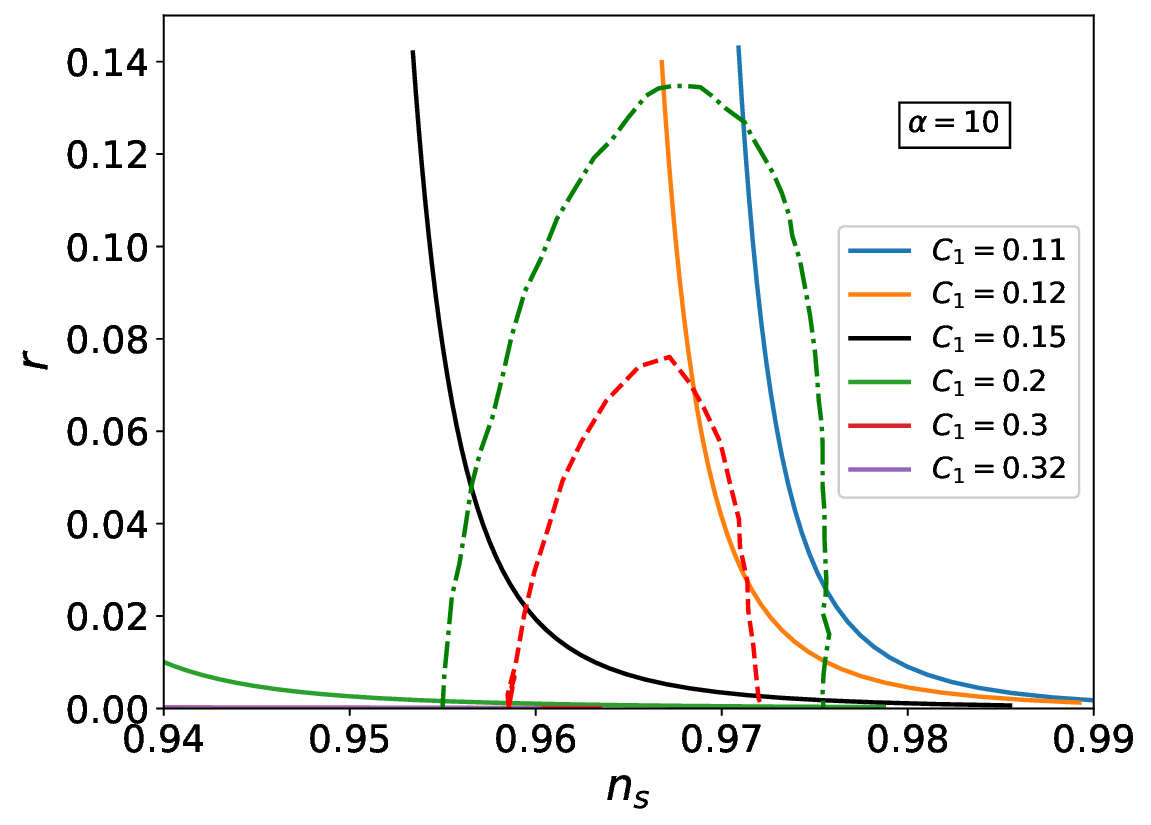}
	\caption{\small Predictions of the strong regime $\alpha$-attractor E-Model with $\alpha = 10$ and plateau-type function $G_1(Q_*)$: Trajectories in the $n_s$--$r$ plane versus Planck 2018 observations.}
	\label{fig-4}
\end{figure*}

\begin{table}[h!]
	\centering
	\begin{tabular}{|c|c|}
		\hline
		$\alpha$ & $C_1$ \\
		\hline
		0.1 & $C_1< 0.001125$ \\
		1 & $C_1< 0.04$ \\
		10 & $C_1< 0.0356$ \\
		100 & $C_1< 0.06324$ \\
		\hline
	\end{tabular}
	\caption{\small Upper bounds on $C_1$ with different values of $\alpha$ (in the allowed range $10^{-2} < \alpha < 10^4$) for the weak regime of $\alpha$-attractor E-model.}
	\label{tab-1}
\end{table}

\subsection{Strong dissipative regime}

Equations~\ref{ns-final} and \ref{r-final} are intrinsically connected through the dissipation ratio $Q_*$, as defined in equation~\ref{Q-star}. This connection implies that, once the functional form of the enhancement function $G(Q_*)$ is specified, the scalar spectral index $n_s$ can be explicitly obtained as a function of the tensor-to-scalar ratio $r$. In this analysis, we consider two physically motivated forms for $G(Q_*)$. The first, given by equation~\ref{plateau}, corresponds to plateau-like potentials frequently encountered in $\alpha$-attractor models. The second, presented in equation~\ref{little}, emerges within the framework of warm little inflation, a scenario inspired by particle physics models that incorporate light degrees of freedom coupled to the inflaton field \cite{Bastero_2016}. By substituting these specific forms of $G(Q_*)$ into our equations, we obtain model-dependent predictions for $n_s$ as a function of $r$, enabling a direct comparison with current observational constraints.

\begin{figure*}
	\includegraphics[scale=0.4]{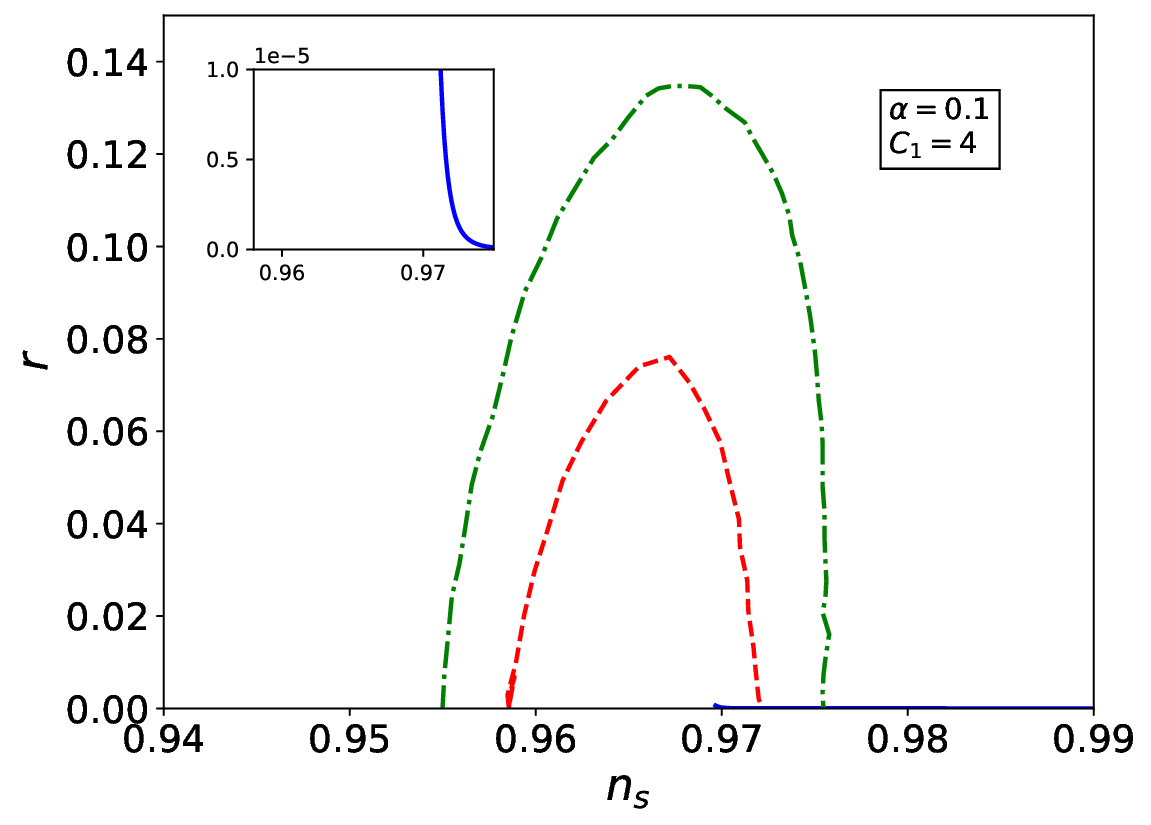}
	\includegraphics[scale=0.4]{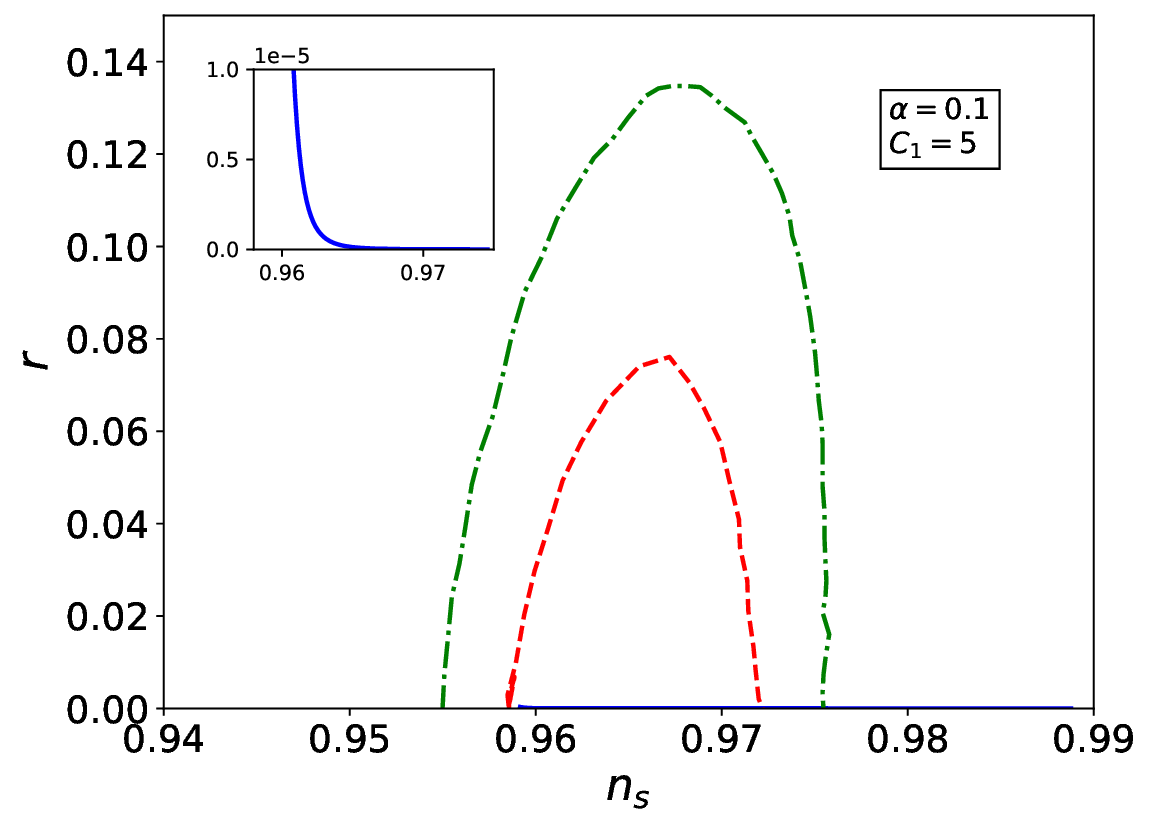}
	\includegraphics[scale=0.4]{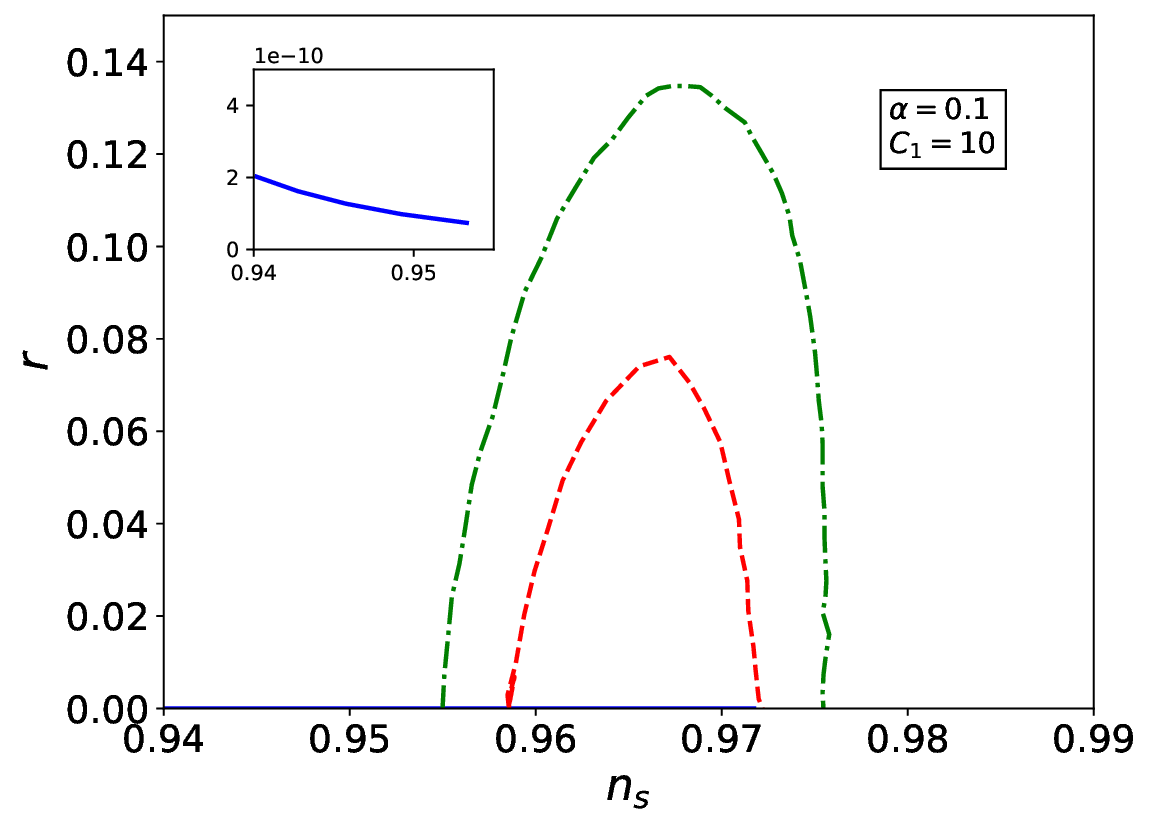}
	\includegraphics[scale=0.4]{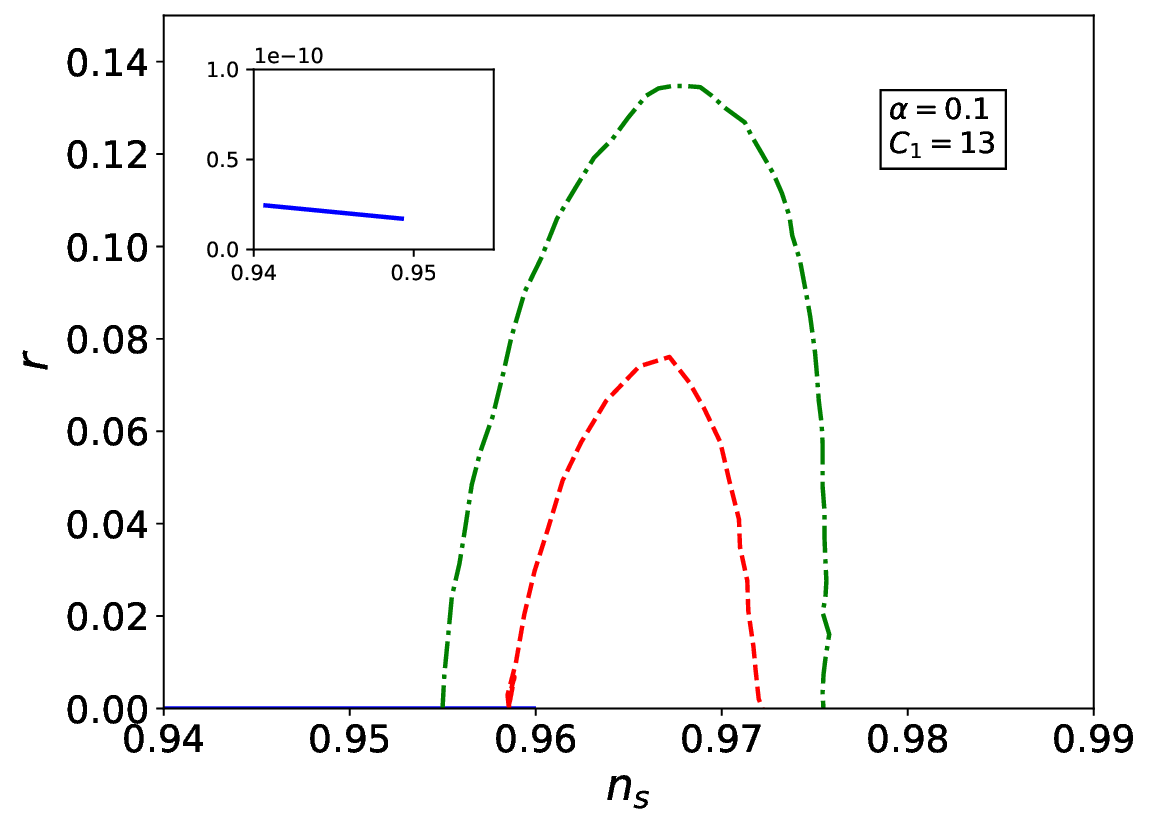}
	\caption{\small Inflationary trajectories in the $n_s$--$r$ plane with different values of $C_1$ for fixed $\alpha=0.1$ (in the admissble range $10^{-2} < \alpha < 10^4$) in the strong regime of $\alpha$-attractor E-model employing the enhancement function $G_2(Q_*)$ for warm little inflation: Comparison with Planck 2018 data.}
	\label{fig-5}
\end{figure*}

Figures~\ref{fig-2}, \ref{fig-3}, and \ref{fig-4} present the evolution of inflationary observables in the $n_s$--$r$ plane with different values of the dissipation parameter $C_1$, corresponding to fixed values of the model parameter $\alpha = 0.1, 1,$ and $10$. These values are representative of the broader interval $10^{-2} < \alpha < 10^4$, which captures the phenomenologically relevant range for $\alpha$-attractor scenarios. The analysis is carried out within the strong dissipative regime of warm inflation, employing the enhancement function $G_1(Q_*)$ derived from plateau-like potentials. Each figure overlays the model predictions with the Planck 2018 confidence contours in the $n_s$--$r$ plane, enabling a clear visual comparison with observational constraints. These plots serve to delineate the allowed parameter space and emphasize the influence of dissipative dynamics on the inflationary predictions of the $\alpha$-attractor E-model.

Table~\ref{tab-2} presents the allowed ranges of the dissipation parameter $C_1$ for various representative values of the model parameter $\alpha$, within the  interval $10^{-2} < \alpha < 10^4$ admissible by Planck 2018. These values cover the theoretically and observationally relevant regime for $\alpha$-attractor models. These results are obtained using the enhancement function $G_1(Q_*)$, which characterizes plateau-like inflationary potentials. The allowed ranges of $C_1$ are determined by requiring consistency between the model's predictions for the scalar spectral index $n_s$ and the tensor-to-scalar ratio $r$, and the observational constraints from the Planck 2018 data. This table provides a concise summary of how the parameter space of warm $\alpha$-attractor E-models is shaped by both the theoretical structure of the enhancement function and current cosmological observations.

\begin{table}[h!]
	\centering
	\begin{tabular}{|c|c|}
		\hline
		$\alpha$ & $C_1$ \\
		\hline
		0.1 & $3.8 < C_1 < 13.5$ \\
		1 & $ 0.6 < C_1 < 0.9$ \\
		10 & $0.12 < C_1 < 0.2$ \\
		\hline
	\end{tabular}
	\caption{\small Allowed ranges of the dissipation parameter $C_1$ with different values of $\alpha$ in the strong dissipative regime of warm inflation, using the enhancement function $G_1(Q_*)$ corresponding to plateau-like potentials.}
	\label{tab-2}
\end{table}

\begin{figure*}
\centering
	\includegraphics[scale=0.5]{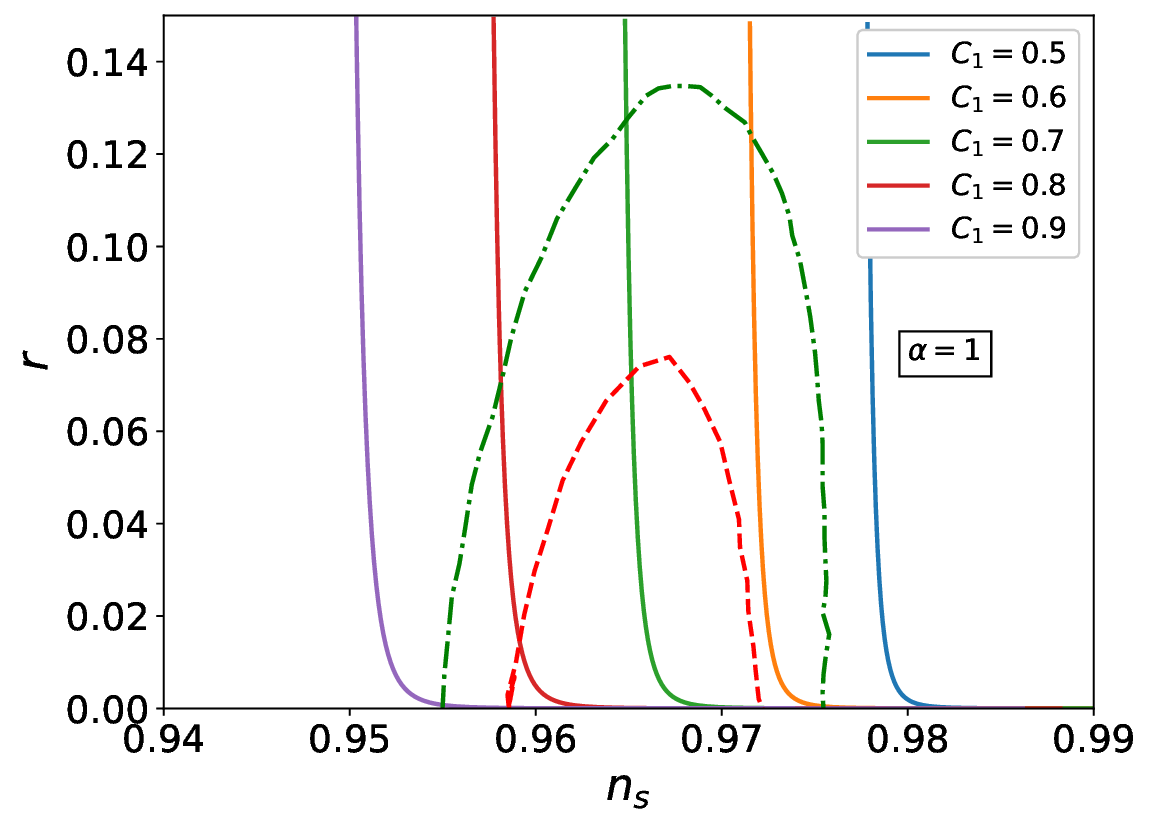}
	\caption{\small Predicted $n_s$–$r$ trajectories with different $C_1$ values in the strong dissipative regime of the $\alpha$-attractor E-Model with $\alpha = 1$, using $G_2(Q_*)$ from warm little inflation, comparison with Planck 2018 constraints.}
	\label{fig-6}
\end{figure*}

\begin{figure*}
\centering
	\includegraphics[scale=0.5]{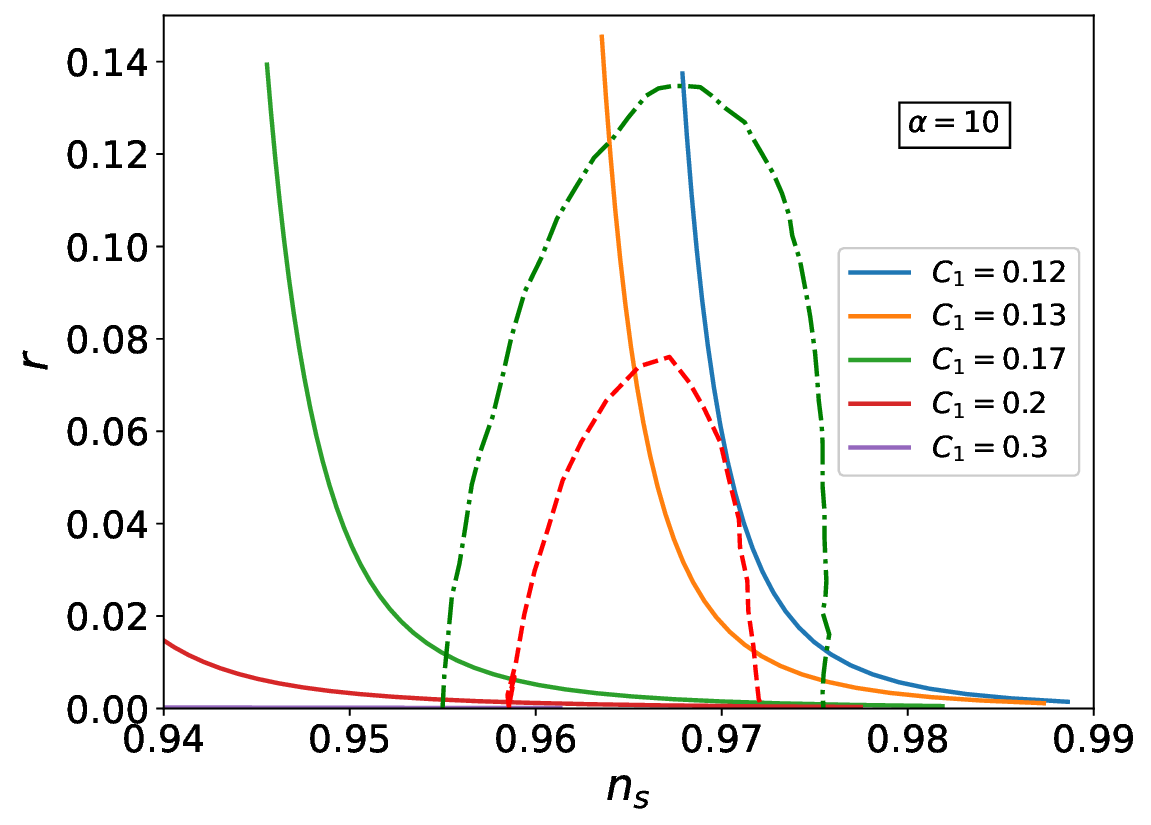}
	\caption{\small Strong regime predictions of the $\alpha$-attractor E-Model with $G_2(Q_*)$ for warm little inflation : Trajectories in the $n_s$--$r$ plane for fixed $\alpha = 10$ and different $C_1$ values, overlaid on Planck 2018 observational bounds.}
	\label{fig-7}
\end{figure*}

Figures~\ref{fig-5}, \ref{fig-6}, and \ref{fig-7} illustrate the predicted trajectories in the $n_s$--$r$ plane for representative values of the dissipation parameter $C_1$, with the model parameter fixed at $\alpha = 0.1, 1, 10$. This value lies within the broader range $10^{-2} < \alpha < 10^4$, which encompasses the regimes allowed by Planck 2018 in the $\alpha$-attractor model. The analysis is performed using the enhancement function $G_2(Q_*)$ associated with the warm little inflation scenario. The resulting trajectories for different values of $C_1$ are plotted alongside the latest Planck 2018 observational contours in the $n_s$--$r$ plane, allowing a direct assessment of the model's consistency with cosmological data. These comparisons serve to constrain the parameter space and highlight the impact of warm dissipative dynamics on inflationary observables in the context of the $\alpha$-attractor E-model.

Table~\ref{tab-3} summarizes the allowed ranges of the dissipation parameter $C_1$ for representative values of the model parameter $\alpha$, within the allowed range $10^{-2} < \alpha < 10^4$. This analysis is conducted utilizing the enhancement function $G_2(Q_*)$, which characterizes the warm little inflation scenario. The entries in Table~\ref{tab-3} are obtained by imposing consistency between the theoretical predictions for the spectral index $n_s$ and the tensor-to-scalar ratio $r$, and the latest observational bounds from Planck 2018. This table provides a clear overview of the viable parameter space for warm little inflation within the $\alpha$-attractor framework, highlighting how the microphysical origin of the dissipation affects the inflationary dynamics and observational viability of the model.

\begin{table}[h!]
	\centering
	\begin{tabular}{|c|c|}
		\hline
		$\alpha$ & $C_1$ \\
		\hline
		0.1 & $4 < C_1 < 13$ \\
		1 & $0.6 < C_1 < 0.08$ \\
		10 & $0.12 < C_1 < 0.3$ \\
		\hline
	\end{tabular}
	\caption{\small Admissible ranges of the dissipation parameter $C_1$ with different values of $\alpha$ in the strong dissipative regime of warm inflation, employing the enhancement function $G_2(Q_*)$ characteristic of warm little inflation.}
	\label{tab-3}
\end{table}

\section{Discussion and Conclusion}
\label{concl}

In this work, we have conducted a detailed analysis of the $\alpha$-attractor E-model within the framework of warm inflation, focusing on both the weak and strong dissipative regimes. Utilizing the latest Planck 2018 observational data, we constrained the model parameters, particularly the dissipation parameter $C_1$ remaining within bounds for the geometric parameter $\alpha$, by examining their effects on the scalar spectral index $n_s$ and the tensor-to-scalar ratio $r$. For the strong dissipative regime, we explored two physically motivated enhancement functions corresponding to plateau-like potentials and warm little inflation scenarios, each capturing distinct microscopic origins of dissipation.

Our results demonstrate that the warm $\alpha$-attractor E-model can produce inflationary observables consistent with Planck 2018 constraints for a viable range of parameter values. Notably, the interplay between $\alpha$ and $C_1$ plays a critical role in shaping the inflationary predictions, with the warm little inflation setup yielding slightly different but compatible constraints compared to the plateau case. These findings reinforce the robustness of the $\alpha$-attractor framework when extended to warm inflation dynamics and highlight the significance of dissipation effects in modifying standard cold inflation predictions.

In conclusion, our study demonstrates that the warm $\alpha$-attractor E-model not only retains the attractive features of its cold counterpart but also gains additional phenomenological flexibility through dissipative effects. These results highlight the inherent strength of the model as a unified framework for describing the early universe, offering both theoretical consistency and compatibility with observational data.


\begin{thebibliography}{10}

\bibitem{guth1981inflationary}
Alan~H. Guth.
\newblock Inflationary universe: A possible solution to the horizon and
  flatness problems.
\newblock {\em Phys. Rev. D}, 23:347--356, Jan 1981.
\newblock URL: \url{https://link.aps.org/doi/10.1103/PhysRevD.23.347}, \href
  {https://doi.org/10.1103/PhysRevD.23.347}
  {\path{doi:10.1103/PhysRevD.23.347}}.

\bibitem{Linde-inflation1982}
Andrei~D. Linde.
\newblock {A New Inflationary Universe Scenario: A Possible Solution of the
  Horizon, Flatness, Homogeneity, Isotropy and Primordial Monopole Problems}.
\newblock {\em Phys. Lett. B}, 108:389--393, 1982.
\newblock \href {https://doi.org/10.1016/0370-2693(82)91219-9}
  {\path{doi:10.1016/0370-2693(82)91219-9}}.

\bibitem{Sato:1980yn}
K.~Sato.
\newblock {First Order Phase Transition of a Vacuum and Expansion of the
  Universe}.
\newblock {\em Mon. Not. Roy. Astron. Soc.}, 195:467--479, 1981.

\bibitem{linde1983chaotic}
Andrei~D. Linde.
\newblock {Chaotic Inflation}.
\newblock {\em Phys. Lett. B}, 129:177--181, 1983.
\newblock \href {https://doi.org/10.1016/0370-2693(83)90837-7}
  {\path{doi:10.1016/0370-2693(83)90837-7}}.

\bibitem{linde1984inflationary}
A~D Linde.
\newblock The inflationary universe.
\newblock {\em Rep. Prog. Phys.}, 47(8):925, aug 1984.
\newblock URL: \url{https://dx.doi.org/10.1088/0034-4885/47/8/002}, \href
  {https://doi.org/10.1088/0034-4885/47/8/002}
  {\path{doi:10.1088/0034-4885/47/8/002}}.

\bibitem{guth1984inflationary}
A.H.Guth and P.J.Steinhardt.
\newblock The inflationary universe.
\newblock {\em Scientific American}, 250:116--129, 1984.
\newblock URL: \url{http://www.jstor.org/stable/24969372}.

\bibitem{Albrecht1982A}
Andreas Albrecht and Paul~J. Steinhardt.
\newblock Cosmology for grand unified theories with radiatively induced
  symmetry breaking.
\newblock {\em Phys. Rev. Lett.}, 48:1220--1223, Apr 1982.
\newblock URL: \url{https://link.aps.org/doi/10.1103/PhysRevLett.48.1220},
  \href {https://doi.org/10.1103/PhysRevLett.48.1220}
  {\path{doi:10.1103/PhysRevLett.48.1220}}.

\bibitem{albrecht1982reheating}
Andreas Albrecht, Paul~J. Steinhardt, Michael~S. Turner, and Frank Wilczek.
\newblock Reheating an inflationary universe.
\newblock {\em Phys. Rev. Lett.}, 48:1437--1440, May 1982.
\newblock URL: \url{https://link.aps.org/doi/10.1103/PhysRevLett.48.1437},
  \href {https://doi.org/10.1103/PhysRevLett.48.1437}
  {\path{doi:10.1103/PhysRevLett.48.1437}}.

\bibitem{HAWKING198235}
S.W. Hawking and I.L. Moss.
\newblock Supercooled phase transitions in the very early universe.
\newblock {\em Phys. Lett. B}, 110(1):35--38, 1982.
\newblock URL:
  \url{https://www.sciencedirect.com/science/article/pii/0370269382909467},
  \href {https://doi.org/https://doi.org/10.1016/0370-2693(82)90946-7}
  {\path{doi:https://doi.org/10.1016/0370-2693(82)90946-7}}.

\bibitem{Larson:2010gs}
D.~Larson et al.
\newblock Seven-year wilkinson microwave anisotropy probe (wmap) observations:
  Power spectra and wmap-derived parameters.
\newblock {\em Astrophys. J. Suppl.}, 192:16, 2011.
\newblock \href {http://arxiv.org/abs/1001.4635} {\path{arXiv:1001.4635}},
  \href {https://doi.org/10.1088/0067-0049/192/2/16}
  {\path{doi:10.1088/0067-0049/192/2/16}}.

\bibitem{Bennett:2010jb}
C.~L. Bennett et al.
\newblock Seven-year wilkinson microwave anisotropy probe (wmap) observations:
  Are there cosmic microwave background anomalies?
\newblock {\em Astrophys. J. Suppl.}, 192:17, 2011.
\newblock \href {http://arxiv.org/abs/1001.4758} {\path{arXiv:1001.4758}},
  \href {https://doi.org/10.1088/0067-0049/192/2/17}
  {\path{doi:10.1088/0067-0049/192/2/17}}.

\bibitem{Jarosik:2010iu}
N.~Jarosik et al.
\newblock Seven-year wilkinson microwave anisotropy probe (wmap) observations:
  Sky maps, systematic errors, and basic results.
\newblock {\em Astrophys. J. Suppl.}, 192:14, 2011.
\newblock \href {http://arxiv.org/abs/1001.4744} {\path{arXiv:1001.4744}},
  \href {https://doi.org/10.1088/0067-0049/192/2/14}
  {\path{doi:10.1088/0067-0049/192/2/14}}.

\bibitem{Hinshaw:2012aka}
G.~Hinshaw et al.
\newblock Nine-year wilkinson microwave anisotropy probe (wmap) observations:
  Cosmological parameter results.
\newblock {\em Astrophys. J. Suppl.}, 208:19, 2013.
\newblock \href {http://arxiv.org/abs/1212.5226} {\path{arXiv:1212.5226}},
  \href {https://doi.org/10.1088/0067-0049/208/2/19}
  {\path{doi:10.1088/0067-0049/208/2/19}}.

\bibitem{Planck:2013pxb}
P.~A.~R. Ade, N.~Aghanim, C.~Armitage-Caplan, M.~Arnaud, M.~Ashdown,
  F.~Atrio-Barandela, and et~al.
\newblock Planck 2013 results. xvi. cosmological parameters.
\newblock {\em Astron. Astrophys.}, 571:A16, 2014.
\newblock \href {http://arxiv.org/abs/1303.5076} {\path{arXiv:1303.5076}},
  \href {https://doi.org/10.1051/0004-6361/201321591}
  {\path{doi:10.1051/0004-6361/201321591}}.

\bibitem{Planck:2015xua}
P.~A.~R. Ade, N.~Aghanim, M.~Arnaud, M.~Ashdown, J.~Aumont, C.~Baccigalupi, and
  et~al.
\newblock {Planck 2015 results. XIII. Cosmological parameters}.
\newblock {\em Astron. Astrophys.}, 594:A13, 2016.
\newblock \href {http://arxiv.org/abs/1502.01589} {\path{arXiv:1502.01589}},
  \href {https://doi.org/10.1051/0004-6361/201525830}
  {\path{doi:10.1051/0004-6361/201525830}}.

\bibitem{Starobinsky:1979}
A.~A. Starobinsky.
\newblock {A new type of isotropic cosmological models without singularity}.
\newblock {\em Phys. Lett. B}, 91:99--102, 1980.
\newblock \href {https://doi.org/10.1016/0370-2693(80)90670-X}
  {\path{doi:10.1016/0370-2693(80)90670-X}}.

\bibitem{Mukhanov:1981}
V.~F. Mukhanov and G.~V. Chibisov.
\newblock {Quantum fluctuations and a nonsingular universe}.
\newblock {\em JETP Lett.}, 33:532--535, 1981.
\newblock [Pisma Zh. Eksp. Teor. Fiz. 33, 549 (1981)].

\bibitem{Hawking:1982}
S.~W. Hawking.
\newblock {The development of irregularities in a single bubble inflationary
  universe}.
\newblock {\em Phys. Lett. B}, 115:295, 1982.
\newblock \href {https://doi.org/10.1016/0370-2693(82)90373-2}
  {\path{doi:10.1016/0370-2693(82)90373-2}}.

\bibitem{Guth:1982}
Alan~H. Guth and So-Young Pi.
\newblock {Fluctuations in the new inflationary universe}.
\newblock {\em Phys. Rev. Lett.}, 49:1110--1113, 1982.
\newblock \href {https://doi.org/10.1103/PhysRevLett.49.1110}
  {\path{doi:10.1103/PhysRevLett.49.1110}}.

\bibitem{Bardeen1983}
James~M. Bardeen, Paul~J. Steinhardt, and Michael~S. Turner.
\newblock Spontaneous creation of almost scale-free density perturbations in an
  inflationary universe.
\newblock {\em Phys. Rev. D}, 28:679--693, Aug 1983.
\newblock URL: \url{https://link.aps.org/doi/10.1103/PhysRevD.28.679}, \href
  {https://doi.org/10.1103/PhysRevD.28.679}
  {\path{doi:10.1103/PhysRevD.28.679}}.

\bibitem{Starobinsky:1982}
A.~A. Starobinsky.
\newblock Dynamics of phase transition in the new inflationary universe
  scenario and generation of perturbations.
\newblock {\em Phys. Lett. B}, 117:175--178, 1982.
\newblock \href {https://doi.org/10.1016/0370-2693(82)90541-X}
  {\path{doi:10.1016/0370-2693(82)90541-X}}.

\bibitem{liddle:1994}
Andrew~R. Liddle, Paul Parsons, and John~D. Barrow.
\newblock Formalizing the slow-roll approximation in inflation.
\newblock {\em Phys. Rev. D}, 50:7222--7232, Dec 1994.
\newblock URL: \url{https://link.aps.org/doi/10.1103/PhysRevD.50.7222}, \href
  {https://doi.org/10.1103/PhysRevD.50.7222}
  {\path{doi:10.1103/PhysRevD.50.7222}}.

\bibitem{LiddleLyth2000}
Andrew~R. Liddle and David~H. Lyth.
\newblock {\em Cosmological Inflation and Large-Scale Structure}.
\newblock Cambridge University Press, New York, 2000.

\bibitem{KolbTurner1990}
Edward~W. Kolb and Michael~S. Turner.
\newblock {\em The Early Universe}, volume~69 of {\em Frontiers in Physics}.
\newblock Addison-Wesley, Redwood City, CA, 1990.

\bibitem{kofman1994reheating}
Lev Kofman, Andrei Linde, and Alexei~A. Starobinsky.
\newblock Reheating after inflation.
\newblock {\em Phys. Rev. Lett.}, 73:3195--3198, Dec 1994.
\newblock URL: \url{https://link.aps.org/doi/10.1103/PhysRevLett.73.3195},
  \href {https://doi.org/10.1103/PhysRevLett.73.3195}
  {\path{doi:10.1103/PhysRevLett.73.3195}}.

\bibitem{kofman1996origin}
Lev~A. Kofman.
\newblock {The Origin of matter in the universe: Reheating after inflation}.
\newblock 5 1996.
\newblock \href {http://arxiv.org/abs/astro-ph/9605155}
  {\path{arXiv:astro-ph/9605155}}.

\bibitem{abbott1982particle}
Laurence~F Abbott, Edward Farhi, and Mark~B Wise.
\newblock Particle production in the new inflationary cosmology.
\newblock {\em Phys. Lett. B}, 117(1-2):29--33, 1982.

\bibitem{allahverdi2010reheating}
Rouzbeh Allahverdi, Robert Brandenberger, Francis-Yan Cyr-Racine, and Anupam
  Mazumdar.
\newblock {Reheating in Inflationary Cosmology: Theory and Applications}.
\newblock {\em Ann. Rev. Nucl. Part. Sci.}, 60:27--51, 2010.
\newblock \href {http://arxiv.org/abs/1001.2600} {\path{arXiv:1001.2600}},
  \href {https://doi.org/10.1146/annurev.nucl.012809.104511}
  {\path{doi:10.1146/annurev.nucl.012809.104511}}.

\bibitem{amin2015nonperturbative}
Mustafa~A. Amin, Mark~P. Hertzberg, David~I. Kaiser, and Johanna Karouby.
\newblock {Nonperturbative Dynamics Of Reheating After Inflation: A Review}.
\newblock {\em Int. J. Mod. Phys. D}, 24:1530003, 2014.
\newblock \href {http://arxiv.org/abs/1410.3808} {\path{arXiv:1410.3808}},
  \href {https://doi.org/10.1142/S0218271815300037}
  {\path{doi:10.1142/S0218271815300037}}.

\bibitem{Berera-1995}
Arjun Berera.
\newblock Warm inflation.
\newblock {\em Phys. Rev. Lett.}, 75:3218--3221, Oct 1995.
\newblock URL: \url{https://link.aps.org/doi/10.1103/PhysRevLett.75.3218},
  \href {https://doi.org/10.1103/PhysRevLett.75.3218}
  {\path{doi:10.1103/PhysRevLett.75.3218}}.

\bibitem{Berera-2001}
Arjun Berera and Rudnei~O. Ramos.
\newblock Affinity for scalar fields to dissipate.
\newblock {\em Phys. Rev. D}, 63:103509, Apr 2001.
\newblock URL: \url{https://link.aps.org/doi/10.1103/PhysRevD.63.103509}, \href
  {https://doi.org/10.1103/PhysRevD.63.103509}
  {\path{doi:10.1103/PhysRevD.63.103509}}.

\bibitem{Berera_2023}
Arjun Berera.
\newblock The warm inflation story.
\newblock {\em Universe}, 9(6), 2023.
\newblock URL: \url{https://www.mdpi.com/2218-1997/9/6/272}, \href
  {https://doi.org/10.3390/universe9060272}
  {\path{doi:10.3390/universe9060272}}.

\bibitem{BERERA2000}
Arjun Berera.
\newblock Warm inflation in the adiabatic regime — a model, an existence
  proof for inflationary dynamics in quantum field theory.
\newblock {\em Nuclear Physics B}, 585(3):666--714, 2000.
\newblock URL:
  \url{https://www.sciencedirect.com/science/article/pii/S0550321300004119},
  \href {https://doi.org/https://doi.org/10.1016/S0550-3213(00)00411-9}
  {\path{doi:https://doi.org/10.1016/S0550-3213(00)00411-9}}.

\bibitem{Bastero_2016}
Mar Bastero-Gil, Arjun Berera, Rudnei~O. Ramos, and Jo\~ao~G. Rosa.
\newblock Warm little inflaton.
\newblock {\em Phys. Rev. Lett.}, 117:151301, Oct 2016.
\newblock URL: \url{https://link.aps.org/doi/10.1103/PhysRevLett.117.151301},
  \href {https://doi.org/10.1103/PhysRevLett.117.151301}
  {\path{doi:10.1103/PhysRevLett.117.151301}}.

\bibitem{Kamali_2015}
V~Kamali and M~R Setare.
\newblock Warm-viscous inflation model on the brane in light of planck data.
\newblock {\em Classical and Quantum Gravity}, 32(23):235005, nov 2015.
\newblock URL: \url{https://dx.doi.org/10.1088/0264-9381/32/23/235005}, \href
  {https://doi.org/10.1088/0264-9381/32/23/235005}
  {\path{doi:10.1088/0264-9381/32/23/235005}}.

\bibitem{Setare2015}
M.~R. Setare and V.~Kamali.
\newblock Warm chaplygin inflation in loop quantum cosmology in light of planck
  data.
\newblock {\em Phys. Rev. D}, 91:123517, Jun 2015.
\newblock URL: \url{https://link.aps.org/doi/10.1103/PhysRevD.91.123517}, \href
  {https://doi.org/10.1103/PhysRevD.91.123517}
  {\path{doi:10.1103/PhysRevD.91.123517}}.

\bibitem{Kamali2016}
Vahid Kamali, Spyros Basilakos, and Ahmad Mehrabi.
\newblock Tachyon warm-intermediate inflation in the light of planck data.
\newblock {\em Eur. Phys. J. C}, 76(10):525, Sep 2016.
\newblock \href {https://doi.org/10.1140/epjc/s10052-016-4380-6}
  {\path{doi:10.1140/epjc/s10052-016-4380-6}}.

\bibitem{Visinelli_2016}
Luca Visinelli.
\newblock Observational constraints on monomial warm inflation.
\newblock {\em Journal of Cosmology and Astroparticle Physics}, 2016(07):054,
  jul 2016.
\newblock URL: \url{https://dx.doi.org/10.1088/1475-7516/2016/07/054}, \href
  {https://doi.org/10.1088/1475-7516/2016/07/054}
  {\path{doi:10.1088/1475-7516/2016/07/054}}.

\bibitem{Motaharfar2016}
Meysam Motaharfar and Hamid~Reza Sepangi.
\newblock Warm-tachyon gauss--bonnet inflation in the light of planck 2015
  data.
\newblock {\em Eur. Phys. J. C}, 76(11):646, Nov 2016.
\newblock \href {https://doi.org/10.1140/epjc/s10052-016-4474-1}
  {\path{doi:10.1140/epjc/s10052-016-4474-1}}.

\bibitem{Panotopoulos2015}
Grigorios Panotopoulos and Nelson Videla.
\newblock Warm $\frac{\lambda}{4}\phi^4$ inflationary universe model in light
  of planck 2015 results.
\newblock {\em Eur. Phys. J. C}, 75(11):525, Nov 2015.
\newblock \href {https://doi.org/10.1140/epjc/s10052-015-3764-3}
  {\path{doi:10.1140/epjc/s10052-015-3764-3}}.

\bibitem{Kamali2018}
Vahid Kamali.
\newblock Non-minimal higgs inflation in the context of warm scenario in the
  light of planck data.
\newblock {\em Eur. Phys. J. C}, 78(11):975, Nov 2018.
\newblock \href {https://doi.org/10.1140/epjc/s10052-018-6449-x}
  {\path{doi:10.1140/epjc/s10052-018-6449-x}}.

\bibitem{Jawad2017}
Abdul Jawad, Shahzad Hussain, Shamaila Rani, and Nelson Videla.
\newblock Impact of generalized dissipative coefficient on warm inflationary
  dynamics in the light of latest planck data.
\newblock {\em Eur. Phys. J. C}, 77(10):700, Oct 2017.
\newblock \href {https://doi.org/10.1140/epjc/s10052-017-5264-0}
  {\path{doi:10.1140/epjc/s10052-017-5264-0}}.

\bibitem{Jawad-2017}
Abdul Jawad, Nelson Videla, and Faiza Gulshan.
\newblock Dynamics of warm power-law plateau inflation with a generalized
  inflaton decay rate: predictions and constraints after planck 2015.
\newblock {\em Eur. Phys. J. C}, 77(5):271, Apr 2017.
\newblock \href {https://doi.org/10.1140/epjc/s10052-017-4846-1}
  {\path{doi:10.1140/epjc/s10052-017-4846-1}}.

\bibitem{Herrera2015}
Ram{\'o}n Herrera, Nelson Videla, and Marco Olivares.
\newblock Warm intermediate inflation in the randall--sundrum ii model in the
  light of planck 2015 and bicep2 results: a general dissipative coefficient.
\newblock {\em Eur. Phys. J. C}, 75(5):205, May 2015.
\newblock \href {https://doi.org/10.1140/epjc/s10052-015-3433-6}
  {\path{doi:10.1140/epjc/s10052-015-3433-6}}.

\bibitem{AlHallak2023}
Mahmoud AlHallak, Khalil~Kalid Al-Said, Nidal Chamoun, and Moustafa~Sayem
  El-Daher.
\newblock On warm natural inflation and planck 2018 constraints.
\newblock {\em Universe}, 9(2), 2023.
\newblock URL: \url{https://www.mdpi.com/2218-1997/9/2/80}, \href
  {https://doi.org/10.3390/universe9020080}
  {\path{doi:10.3390/universe9020080}}.

\bibitem{Benetti2017}
Micol Benetti and Rudnei~O. Ramos.
\newblock Warm inflation dissipative effects: Predictions and constraints from
  the planck data.
\newblock {\em Phys. Rev. D}, 95:023517, Jan 2017.
\newblock URL: \url{https://link.aps.org/doi/10.1103/PhysRevD.95.023517}, \href
  {https://doi.org/10.1103/PhysRevD.95.023517}
  {\path{doi:10.1103/PhysRevD.95.023517}}.

\bibitem{Kallosh:2013}
Renata Kallosh and Andrei Linde.
\newblock {Universality Class in Conformal Inflation}.
\newblock {\em JCAP}, 1307:002, 2013.
\newblock \href {http://arxiv.org/abs/1306.5220} {\path{arXiv:1306.5220}},
  \href {https://doi.org/10.1088/1475-7516/2013/07/002}
  {\path{doi:10.1088/1475-7516/2013/07/002}}.

\bibitem{Roest:2014}
Diederik Roest.
\newblock {Universality classes of inflation}.
\newblock {\em JCAP}, 1401:007, 2014.
\newblock \href {http://arxiv.org/abs/1309.1285} {\path{arXiv:1309.1285}},
  \href {https://doi.org/10.1088/1475-7516/2014/01/007}
  {\path{doi:10.1088/1475-7516/2014/01/007}}.

\bibitem{Bastero-Gil_2013}
Mar Bastero-Gil, Arjun Berera, Rudnei~O. Ramos, and João~G. Rosa.
\newblock General dissipation coefficient in low-temperature warm inflation.
\newblock {\em Journal of Cosmology and Astroparticle Physics}, 2013(01):016,
  jan 2013.
\newblock URL: \url{https://dx.doi.org/10.1088/1475-7516/2013/01/016}, \href
  {https://doi.org/10.1088/1475-7516/2013/01/016}
  {\path{doi:10.1088/1475-7516/2013/01/016}}.

\bibitem{Zhang_2009}
Yi~Zhang.
\newblock Warm inflation with a general form of the dissipative coefficient.
\newblock {\em Journal of Cosmology and Astroparticle Physics}, 2009(03):023,
  mar 2009.
\newblock URL: \url{https://dx.doi.org/10.1088/1475-7516/2009/03/023}, \href
  {https://doi.org/10.1088/1475-7516/2009/03/023}
  {\path{doi:10.1088/1475-7516/2009/03/023}}.

\bibitem{Planck2018}
Y.~Akrami et~al.
\newblock {Planck 2018 results. X. Constraints on inflation}.
\newblock {\em Astron. Astrophys.}, 641:A10, 2020.
\newblock \href {http://arxiv.org/abs/1807.06211} {\path{arXiv:1807.06211}},
  \href {https://doi.org/10.1051/0004-6361/201833887}
  {\path{doi:10.1051/0004-6361/201833887}}.

\bibitem{PRL-Planck}
P.~A.~R. Ade~et al.
\newblock Improved constraints on primordial gravitational waves using planck,
  wmap, and bicep/keck observations through the 2018 observing season.
\newblock {\em Phys. Rev. Lett.}, 127:151301, Oct 2021.
\newblock URL: \url{https://link.aps.org/doi/10.1103/PhysRevLett.127.151301},
  \href {https://doi.org/10.1103/PhysRevLett.127.151301}
  {\path{doi:10.1103/PhysRevLett.127.151301}}.

\bibitem{Kallosh2013}
Renata Kallosh, Andrei Linde, and Diederik Roest.
\newblock Superconformal inflationary $\alpha$-attractors.
\newblock {\em JHEP}, 11:198, 2013.
\newblock \href {http://arxiv.org/abs/1311.0472} {\path{arXiv:1311.0472}},
  \href {https://doi.org/10.1007/JHEP11(2013)198}
  {\path{doi:10.1007/JHEP11(2013)198}}.

\bibitem{Nozari:2018}
Kourosh Nozari and Narges Rashidi.
\newblock Observational viability of an inflation model with e-model nonminimal
  derivative coupling.
\newblock {\em The Astrophysical Journal}, 863(2), 2018.
\newblock \href {https://doi.org/10.3847/1538-4357/aad18e}
  {\path{doi:10.3847/1538-4357/aad18e}}.

\bibitem{Salamate:2019}
F.~Salamate, H.~Sheikhahmadi, and K.~Saaidi.
\newblock E-model $\alpha$ -attractor on brane from planck data and reheating
  temperature.
\newblock {\em Journal of Experimental and Theoretical Physics}, 129:774--780,
  2019.
\newblock \href {https://doi.org/10.3103/S0027134919050114}
  {\path{doi:10.3103/S0027134919050114}}.

\bibitem{Kallosh2013b}
Renata Kallosh and Andrei Linde.
\newblock Superconformal generalizations of the starobinsky model.
\newblock {\em Journal of Cosmology and Astroparticle Physics}, 2013(06):028,
  2013.
\newblock URL:
  \url{https://iopscience.iop.org/article/10.1088/1475-7516/2013/06/028}, \href
  {https://doi.org/10.1088/1475-7516/2013/06/028}
  {\path{doi:10.1088/1475-7516/2013/06/028}}.

\bibitem{Galante:2014}
Mario Galante, Renata Kallosh, Andrei Linde, and Diederik Roest.
\newblock {Unity of Cosmological Inflation Attractors}.
\newblock {\em Phys. Rev. Lett.}, 114:141302, 2015.
\newblock \href {http://arxiv.org/abs/1412.3797} {\path{arXiv:1412.3797}},
  \href {https://doi.org/10.1103/PhysRevLett.114.141302}
  {\path{doi:10.1103/PhysRevLett.114.141302}}.

\bibitem{Kallosh:2015zsa}
Renata Kallosh and Andrei Linde.
\newblock {Planck, LHC, and $\alpha$-attractors}.
\newblock {\em Phys. Rev. D}, 91(8):083528, 2015.
\newblock \href {http://arxiv.org/abs/1502.07733} {\path{arXiv:1502.07733}},
  \href {https://doi.org/10.1103/PhysRevD.91.083528}
  {\path{doi:10.1103/PhysRevD.91.083528}}.

\bibitem{MOSS1985}
I.G. Moss.
\newblock Primordial inflation with spontaneous symmetry breaking.
\newblock {\em Phys. Lett. B}, 154(2):120--124, 1985.
\newblock URL:
  \url{https://www.sciencedirect.com/science/article/pii/0370269385905702},
  \href {https://doi.org/https://doi.org/10.1016/0370-2693(85)90570-2}
  {\path{doi:https://doi.org/10.1016/0370-2693(85)90570-2}}.

\bibitem{Berera_1996}
Arjun Berera.
\newblock Thermal properties of an inflationary universe.
\newblock {\em Phys. Rev. D}, 54:2519--2534, Aug 1996.
\newblock URL: \url{https://link.aps.org/doi/10.1103/PhysRevD.54.2519}, \href
  {https://doi.org/10.1103/PhysRevD.54.2519}
  {\path{doi:10.1103/PhysRevD.54.2519}}.

\bibitem{Berera_Fluc_1995}
Arjun Berera and Li-Zhi Fang.
\newblock Thermally induced density perturbations in the inflation era.
\newblock {\em Phys. Rev. Lett.}, 74:1912--1915, Mar 1995.
\newblock URL: \url{https://link.aps.org/doi/10.1103/PhysRevLett.74.1912},
  \href {https://doi.org/10.1103/PhysRevLett.74.1912}
  {\path{doi:10.1103/PhysRevLett.74.1912}}.

\bibitem{Berera:1996fm}
A.~Berera.
\newblock {Interpolating the Stage of Inflationary Universe: From Slow-roll to
  Warm Inflation}.
\newblock {\em Phys. Rev. D}, 54:2519--2534, 1996.
\newblock \href {http://arxiv.org/abs/hep-th/9601134}
  {\path{arXiv:hep-th/9601134}}, \href
  {https://doi.org/10.1103/PhysRevD.54.2519}
  {\path{doi:10.1103/PhysRevD.54.2519}}.

\bibitem{Berera:1999ws}
A.~Berera.
\newblock {Warm Inflation at Arbitrary Adiabaticity: A Model, an Existence
  Proof for Inflationary Dynamics in Quantum Field Theory}.
\newblock {\em Nucl. Phys. B}, 585:666--714, 2000.
\newblock \href {http://arxiv.org/abs/hep-ph/9904409}
  {\path{arXiv:hep-ph/9904409}}, \href
  {https://doi.org/10.1016/S0550-3213(00)00411-9}
  {\path{doi:10.1016/S0550-3213(00)00411-9}}.

\bibitem{Hall_2004}
Lisa M.~H. Hall, Ian~G. Moss, and Arjun Berera.
\newblock Scalar perturbation spectra from warm inflation.
\newblock {\em Phys. Rev. D}, 69:083525, Apr 2004.
\newblock URL: \url{https://link.aps.org/doi/10.1103/PhysRevD.69.083525}, \href
  {https://doi.org/10.1103/PhysRevD.69.083525}
  {\path{doi:10.1103/PhysRevD.69.083525}}.

\bibitem{Moss_2008}
Ian~G Moss and Chun Xiong.
\newblock On the consistency of warm inflation.
\newblock {\em Journal of Cosmology and Astroparticle Physics}, 2008(11):023,
  nov 2008.
\newblock URL: \url{https://dx.doi.org/10.1088/1475-7516/2008/11/023}, \href
  {https://doi.org/10.1088/1475-7516/2008/11/023}
  {\path{doi:10.1088/1475-7516/2008/11/023}}.

\bibitem{bastero2018}
Mar Bastero-Gil, Arjun Berera, Rafael Hern\'andez-Jim\'enez, and Jo\~ao~G.
  Rosa.
\newblock Dynamical and observational constraints on the warm little inflaton
  scenario.
\newblock {\em Phys. Rev. D}, 98:083502, Oct 2018.
\newblock URL: \url{https://link.aps.org/doi/10.1103/PhysRevD.98.083502}, \href
  {https://doi.org/10.1103/PhysRevD.98.083502}
  {\path{doi:10.1103/PhysRevD.98.083502}}.

\bibitem{Graham:2009}
Chris Graham and Ian~G. Moss.
\newblock {Density fluctuations from warm inflation}.
\newblock {\em JCAP}, 0907:013, 2009.
\newblock \href {http://arxiv.org/abs/0905.3500} {\path{arXiv:0905.3500}},
  \href {https://doi.org/10.1088/1475-7516/2009/07/013}
  {\path{doi:10.1088/1475-7516/2009/07/013}}.

\bibitem{Bastero-Gil:2009-model}
Mar Bastero-Gil and Arjun Berera.
\newblock {Warm inflation model building}.
\newblock {\em Int. J. Mod. Phys. A}, 24:2207--2240, 2009.
\newblock \href {http://arxiv.org/abs/0902.0521} {\path{arXiv:0902.0521}},
  \href {https://doi.org/10.1142/S0217751X09044206}
  {\path{doi:10.1142/S0217751X09044206}}.

\bibitem{Ramos:2013}
Rudnei~O. Ramos and L.~A. da~Silva.
\newblock {Power spectrum for inflation models with quantum and thermal
  noises}.
\newblock {\em JCAP}, 1303:032, 2013.
\newblock \href {http://arxiv.org/abs/1302.3544} {\path{arXiv:1302.3544}},
  \href {https://doi.org/10.1088/1475-7516/2013/03/032}
  {\path{doi:10.1088/1475-7516/2013/03/032}}.

\bibitem{Taylor:2000}
A.~N. Taylor and Arjun Berera.
\newblock {Perturbation spectra in the warm inflationary scenario}.
\newblock {\em Phys. Rev. D}, 62:083517, 2000.
\newblock \href {http://arxiv.org/abs/astro-ph/0006077}
  {\path{arXiv:astro-ph/0006077}}, \href
  {https://doi.org/10.1103/PhysRevD.62.083517}
  {\path{doi:10.1103/PhysRevD.62.083517}}.

\bibitem{deOliveira:2001}
H.~P. de~Oliveira and S.~E. Joras.
\newblock {On perturbations in warm inflation}.
\newblock {\em Phys. Rev. D}, 64:063513, 2001.
\newblock \href {http://arxiv.org/abs/gr-qc/0103089}
  {\path{arXiv:gr-qc/0103089}}, \href
  {https://doi.org/10.1103/PhysRevD.64.063513}
  {\path{doi:10.1103/PhysRevD.64.063513}}.

\bibitem{Bezrukov_2009}
F.~Bezrukov, D.~Gorbunov, and M.~Shaposhnikov.
\newblock On initial conditions for the hot big bang.
\newblock {\em J. Cosmol. Astropart. Phys.}, 2009(06):029,
  jun 2009.
\newblock URL: \url{https://dx.doi.org/10.1088/1475-7516/2009/06/029}, \href
  {https://doi.org/10.1088/1475-7516/2009/06/029}
  {\path{doi:10.1088/1475-7516/2009/06/029}}.

\bibitem{BASTERO-2009}
Mar Bastero-Gil and Arjun Berera.
\newblock Warm inflation model building.
\newblock {\em Int. J. Mod. Phys. A}, 24(12):2207--2240,
  2009.
\newblock \href
  {http://arxiv.org/abs/https://doi.org/10.1142/S0217751X09044206}
  {\path{arXiv:https://doi.org/10.1142/S0217751X09044206}}, \href
  {https://doi.org/10.1142/S0217751X09044206}
  {\path{doi:10.1142/S0217751X09044206}}.

\bibitem{aghanim2020planck}
N~Aghanim, Y~Akrami, M~Ashdown, J~Aumont, C~Baccigalupi, M~Ballardini, A.J
  Banday, R.B Barreiro, N~Bartolo, S~Basak, et~al.
\newblock Planck 2018 results-VI. Cosmological parameters.
\newblock {\em Astron. Astrophys.}, 641:A6, 2020.

\end{thebibliography}

\end{document}